\documentclass[useAMS,usenatbib,english,a4paper,twocolumn,fleqn]{mnras}
\pdfoutput=1

\usepackage{newtxtext,newtxmath}
\usepackage[T1]{fontenc}
\usepackage{ae,aecompl}

\usepackage{graphicx}	
\usepackage{amsmath}	
\usepackage{amssymb}	
\usepackage{subfig}      
\usepackage{natbib}
\usepackage{booktabs}
\usepackage{cleveref}
\crefname{figure}{Fig.}{Figs.}
\crefname{table}{Table}{Tables}
\crefname{chapter}{Chapter}{Chapters}

\newcommand{\perc}[1]{$#1$ per cent}
\newcommand{\units}[1]{\, \mathrm{#1}}
\newcommand{\unitstx}[1]{\mathrm{#1}}
\newcommand{\diff}{\mathop{}\!\mathrm{d}}

\newcommand{\pderiv}[2]{\frac{\mathrm{\partial} #1}{\mathrm{\partial} {#2}}}
\newcommand{\rfsec}[1]{\mbox{\S\ref{sec:#1}}}
\newcommand{\equnp}[1]{eq.~\ref{eq:#1}}

\newcommand{\subrf}[1]{\protect\subref{#1}}
\newcommand{\subrfig}[1]{\protect\subref{fig:#1}}
\newcommand{\vir}[1]{#1_{\mathrm{vir} }}

\newcommand{\mstar}{$M_\ast$}
\newcommand{\mach}{\mathcal{M}}
\newcommand{\kboltz}{k_{\mathrm{B}}}
\newcommand{\zeq}[1]{\mbox{$z=#1$}}
\newcommand{\nbody}{$N$-body~}
\newcommand{\delvir}{\Delta_{\mathrm{vir}}}
\newcommand{\msun}{\units{M_\odot}}
\newcommand{\Rv}{R_{\mathrm{vir}}}
\newcommand{\Mv}{M_{\mathrm{vir}}}
\newcommand{\Vv}{V_{\mathrm{vir}}}
\newcommand{\Tv}{T_{\mathrm{vir}}}

\graphicspath{ {figs/} }
\DeclareGraphicsExtensions{.pdf,.eps}

\title[CFs \& Shocks Formed by Streams in Galaxy Clusters]{Cold Fronts and Shocks Formed by Gas Streams in Galaxy Clusters}

\author[Zinger et al.]{E.\@ Zinger$^1$\thanks{E-mail: elad.zinger@mail.huji.ac.il}, A.\@ Dekel$^1$, Y.\@ Birnboim$^{1}$, D.\@ Nagai$^2$, E.\@ Lau$^2$ \&  A.\@ V.\@ Kravtsov$^3$\\
  $^1$Center for Astrophysics and Planetary Science, Racah Institute of Physics, The Hebrew University, Jerusalem 91904, Israel\\
  $^2$Department of Physics, Yale University, New Haven, CT 06520, USA\\
  $^3$Department of Astronomy \& Astrophysics, The University of Chicago, Chicago, IL 60637 USA }

\date{Accepted 2018 January12. Received January 12; in original form 2016
  September 16}

\pubyear{2018}

\begin{document}
\pagerange{\pageref{firstpage}--\pageref{lastpage}}
\maketitle
\label{firstpage}

\begin{abstract}
Cold Fronts and shocks are hallmarks of the complex intra-cluster
medium (ICM) in galaxy clusters. They are thought to occur due to gas
motions within the ICM and are often attributed to galaxy mergers
within the cluster. Using hydro-cosmological simulations of clusters
of galaxies, we show that collisions of inflowing gas streams, seen to
penetrate to the very centre of about half the clusters, offer an
additional mechanism for the formation of shocks and cold fronts in
cluster cores. Unlike episodic merger events, a gas stream inflow
persists over a period of several $\unitstx{Gyrs}$ and it could
generate a particular pattern of multiple cold fronts and shocks.
\end{abstract}


\section{Introduction}\label{sec:intro}
X-ray observations of the gaseous Intra-Cluster Medium (ICM) reveal it
is rife with features including merging substructures, cavities, shock
waves and Cold Fronts (CF) \citep{Markevitch2007}.

The Cold Fronts are contact discontinuities of constant or smoothly
changing pressure and velocity over a sharp interface between two
regions in which one is denser and cooler than its neighbor. A
discontinuous drop in temperature across the interface is paired with
a jump in density. If no significant non-thermal pressure components are
present, the temperature and density contrasts across the interface
will be inversely equal.

Shocks on the other hand are characterized by discontinuous jumps in
the pressure, density and temperature, all of which increase in value
across the shock front. The two phenomena are connected since many
processes by which a shock is formed also entail the formation of a
contact discontinuity behind it.

CFs are very common in clusters (see \citealt{Markevitch2007} for a
comprehensive review) and have been found in a variety of sizes and
shapes (e.g.\@ concentric arcs, filaments, radial or spiral) both in
observations \citep{Ghizzardi2010} and simulations
\citep{Bialek2002,Nagai2003,Poole2006,Hallman2010}, and are found in
all environments (e.g.\@ disturbed versus quiescent). The accepted
measure of a CF strength is the density or temperature contrast which
is commonly observed to be scattered about a value of $\simeq 2$
\citep{Owers2009}. In many cases a jump in gas metallicity is also
observed, possibly suggesting that enriched low entropy gas stripped
from satellite galaxies in the cluster is at play
\citep{Markevitch2000}. Signatures of substantial shear flows in CF
were detected in relaxed cores \citep{Keshet2010b}.  \citet{Reiss2014}
found evidence for strong magnetic fields parallel to the CF, which
may contribute to their stability.

Several different mechanisms have been put forward to explain the
origin of CFs. They may form as the interface between the low entropy
gas left over from merging satellites and the hot diffuse ICM
\citep{Markevitch2000}. Other mechanisms invoke shocks, since contact
discontinuities are often found behind shocks. Many processes which
can induce shocks in the ICM will lead to the formation of CFs:
merging substructure \citep{Nagai2003,Owers2011}, `gas sloshing' about
the centre of the potential well of the cluster due to mergers
\citep{Churazov2003,Ascasibar2006,ZuHone2010,ZuHone2013} or
oscillations of the dark Matter distribution \citep{Tittley2005}.

Processes which produce local instabilities can also lead to the
formation of CF in the ICM. Local thermal instabilities can lead to
the condensations of cold gas, even when the ICM is globally stable
\citep{Sharma2012,Gaspari2012,Li2015,Prasad2015}. \citet{Balbus2010}
speculate that the non-linear evolution of over-stable states which
occur when radiative and thermal processes act to stabilize the
heat-flux-driven buoyancy instability \citep[HBI,][]{Parrish2008a} and
the magneto-thermal instability \citep[MTI,][]{Parrish2008} can in
turn lead to the formation of CF. The mechanisms cited above are
noteworthy in that they offer an explanation for the existence of CFs
in relaxed clusters which are thought to be devoid of dynamically
violent processes.

A merger of two cluster sized systems can generate myriad structures
and features in the ICM of the merger remnant. In \citet{Poole2006}, a
survey of simulated merger events is carried out and shows that CFs of
different forms, as well as other transient morphological features
(`bridges', `bubbles', `edges', etc.\@) can be generated in these
cataclysmic events.

\citet{Birnboim2010} showed that when trailing shocks merge a CF is
formed and thus co-centric CFs in the ICM can be the result of shocks
which originated at the centre and merged with the virial accretion
shock in the past.

In this paper we suggest yet another mechanism for generating shocks
and CFs in the ICM, related to the smooth accretion of mass
along filaments into the cluster. In the past decade, the
issue of gas accretion on to galaxies and clusters has been
overhauled, with the idealized spherical infall scenario
\citep{White1978} being replaced by accretion that occurs
predominately along filamentary streams which flow along the
large-scale dark matter cosmic web
\citep{Birnboim2003,Keres2005,Dekel2006,Dekel2009,Keres2009}.

Two modes of gas accretion are identified in cosmological simulations
in galactic haloes at high redshift: hot gas which has been
shock-heated by the virial shock and accretes spherically via cooling
to the centre and cold gas which accretes through dense filaments
originating in the cosmic web which, due to their higher
density and shorter cooling times, are impervious to the formation of
a shock \citep{Birnboim2003}. The cold mode accretion is dominant for
low-mass haloes while the hot mode accretion becomes more important
for high-mass haloes \citep{Dekel2006}.

The new understanding that gas streams are an important feature in the
formation of galaxies and clusters, especially in terms of mass
accretion into the system, may have far reaching implications on the
way cluster-sized systems are formed and maintained.

\citet{Dekel2006} and \citet{Dekel2009} show that the coexistence of the hot
and cold modes of accretion reflects the interplay between the shock-heating
scale and the dark matter non-linear clustering scale (the Press-Schechter
mass), \mstar. In standard cosmologies, the large-scale structure of the dark
matter is roughly self-similar. Haloes of mass \mbox{$\sim M_{\ast}$} are
embedded within the filaments and as a result, mass infall will be scattered
over a wide solid angle. In the much rarer haloes of \mbox{$M\gg M_{\ast}$},
and galaxy clusters fall firmly into this category, the accretion will be
predominately along filaments which are thin in comparison to the halo size,
and significantly denser than their host haloes. For massive filaments, more
massive than $10^{12}\msun/\units{Mpc}$, a shock is expected to form at the
edges of filaments which feed low-redshift clusters with $M_{halo}\gtrsim
10^{15}\msun$ \citep{Birnboim2016}. Thus, gas which is accreted along
filaments on to clusters is pre-heated to $\gtrsim 10^6\units{K}$ prior to its
entry to the cluster virial radius.

In \citet{Zinger2016}, a study of a suite of simulated clusters demonstrated
that the filamentary streams which flow from the cosmic web into the cluster,
are still found in clusters at \zeq{0} and are still the channel in which most
of the mass accretion into the cluster takes places. In about half of these
clusters, the streams were found to penetrate into the inner regions of the
cluster to within 25 per cent of the cluster virial radius, and often even
within $0.1\Rv$. The gas in these streams, already at $\gtrsim 10^6\units{K}$,
is further heated as it flows towards the centre.  By the time it reaches the
inner regions of the cluster, the gas stream is at the virial temperature and
no longer cooler than the ambient gas.

Of the clusters examined by \citet{Zinger2016}, in which the streams penetrate
into the very centre of the cluster, all were independently classified as
dynamically `unrelaxed'. Conversely, in the `relaxed' clusters examined, the
streams did not penetrate deeper than $\gtrsim 0.35 \Rv$. Thus the dynamical
state of the cluster was shown to be linked with the presence or absence of
deeply penetrating streams in the central regions, such that a deeply
penetrating stream can lead to an unrelaxed cluster. In addition, it was found
that the degree of penetration of the streams in a given cluster can change
over the evolution of the cluster over typical timescales of
$\sim\unitstx{Gyr}$, with the dynamical state changing accordingly.

The purpose of this paper is to demonstrate that the inflowing gas
streams in clusters can generate shocks and CFs in the ICM, and to
this end we have chosen to focus with some detail on three cluster
outputs, where this phenomenon is convincingly exhibited.

The paper is organized as follows. In \rfsec{coldFronts}, we treat the
case of gas stream collision and CF formation via a simple analytic
model. In \rfsec{sims}, we describe the simulations used for the
analysis and in \rfsec{results} we present three representative
examples of gas streams colliding in simulated clusters and forming
shocks and CFs. In \rfsec{detect} we examine the potential for
observationally detecting the CFs and shocks in our examples and in
\rfsec{discuss} we summarize and discuss our findings.

\begin{figure}
  \centering
  \subfloat[Instance of collision]{\label{fig:collision_initial}
    \includegraphics[width=8cm,keepaspectratio,bb=0 0 9in 5.01in]{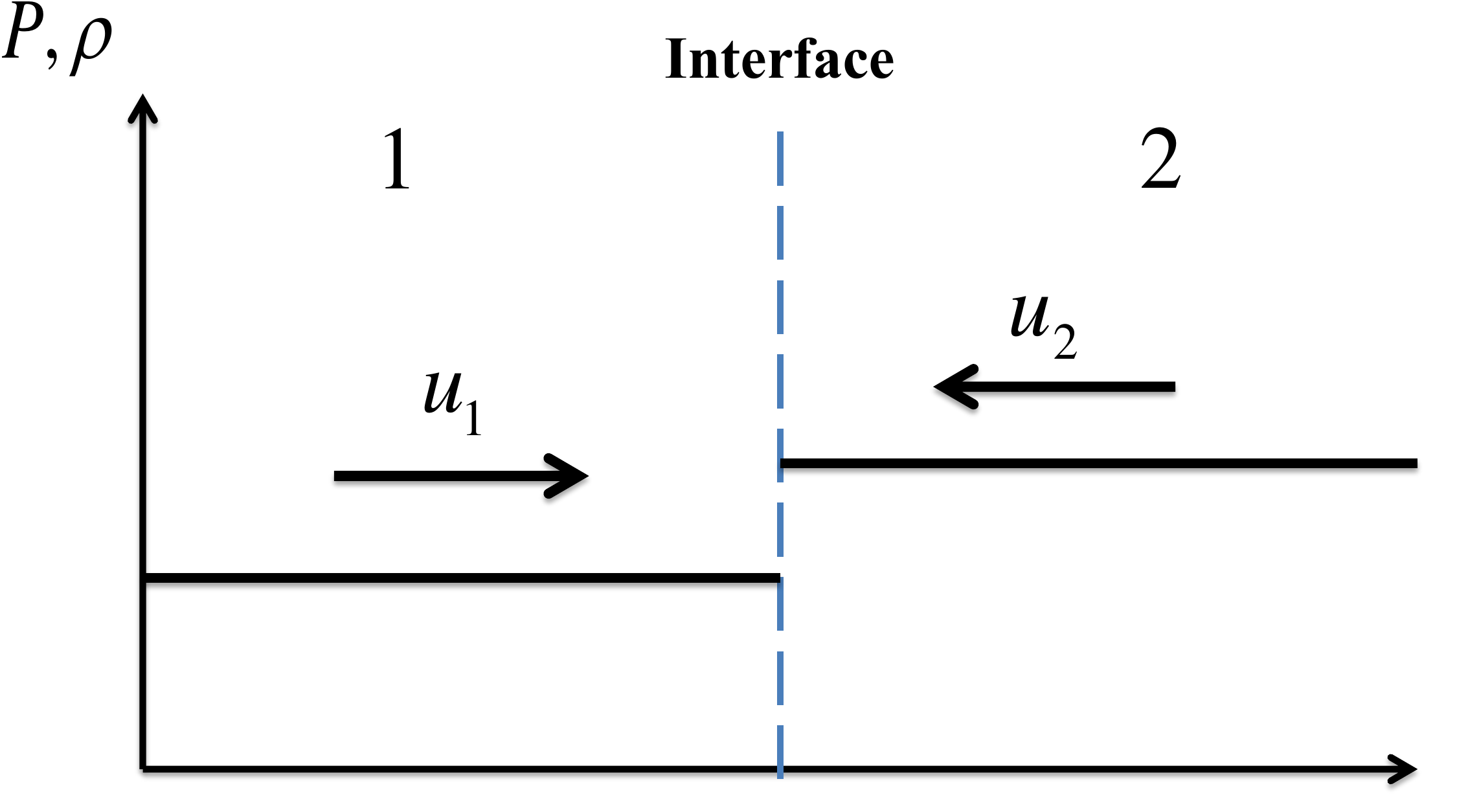}}\\
  \subfloat[Pressure at a later stage]{\label{fig:collisionPressure_later}
    \includegraphics[width=8cm,keepaspectratio,bb=0 0 10in 7.5in,trim=0.4in 1.30in 0.85in  1.20in, clip]{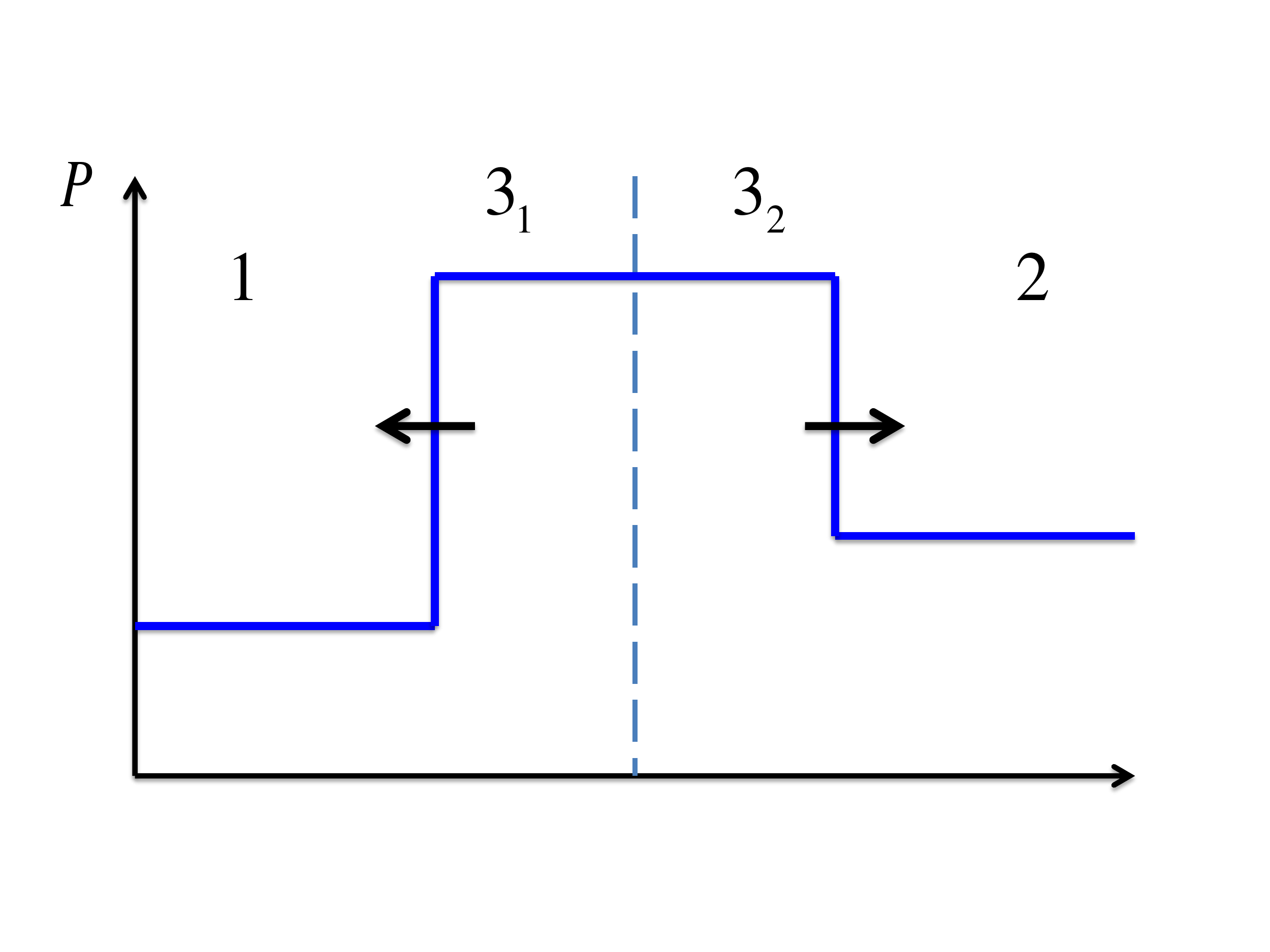}}\\
  \subfloat[Density at a later stage]{\label{fig:collisionDensity_later}
    \includegraphics[width=8cm,keepaspectratio,bb=0 0 10in 7.5in,trim=0.4in 1.30in 0.85in  1.20in, clip]{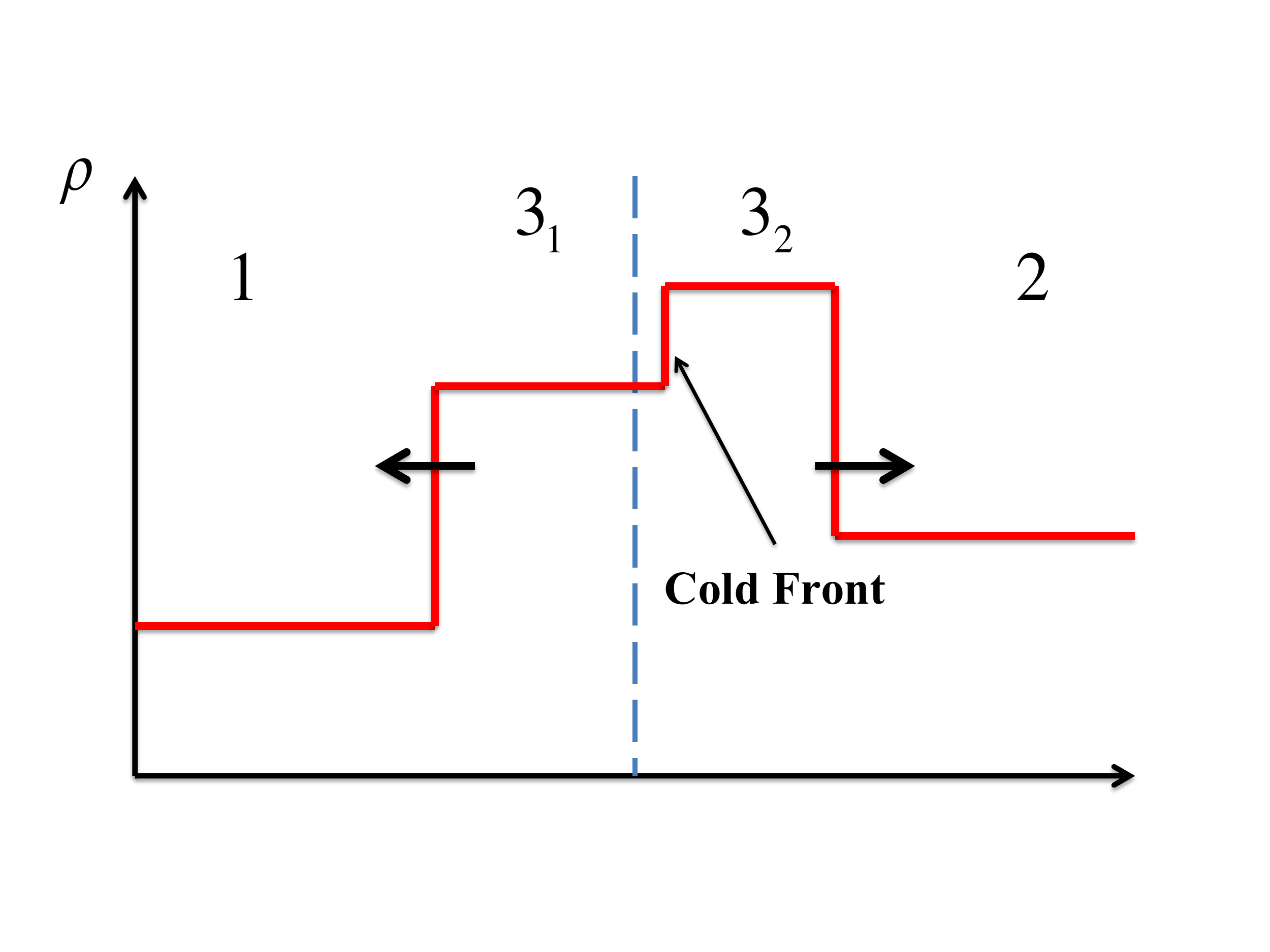}}
  \caption{ Schematic representation of the collision between two
    homogeneous media. In the instance of collision
    \subrf{fig:collision_initial}, two media of constant pressure and
    density are colliding. At a later time, two shocks have formed and
    are propagating in opposite directions. Between the two shocks a
    third post-shock state of constant pressure and velocity has
    formed \subrf{fig:collisionPressure_later}. The density in the
    post shock \subrf{fig:collisionDensity_later} region has one of
    two values (set by the jump conditions for each shock wave)
    bridged by a contact discontinuity which forms at the Lagrangian
    location of the initial collision site and moves at the post-shock
    velocity.}
  \label{fig:collision}
\end{figure}
\begin{figure}
  \centering
  \includegraphics[width=8.5cm,keepaspectratio,bb=0 0 5.4in 4.32in]{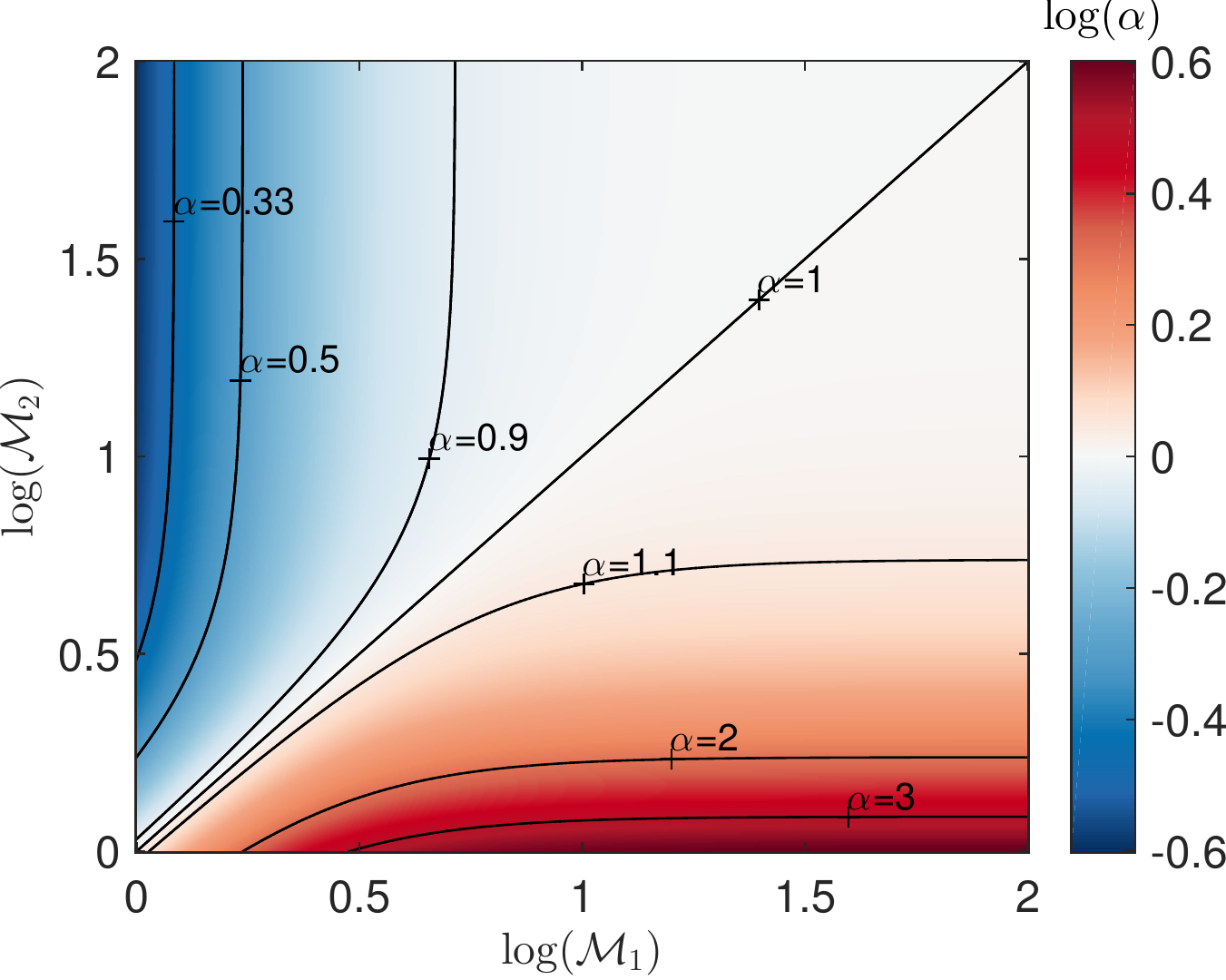}
  \caption{The value of the ratio of the post-shock density contrast
    to the initial density contrast $\alpha=q/q_0$ in an idealized
    stream collision scenario for a range of Mach numbers for the two
    shocks, for an ideal mono-atomic gas with $\gamma=5/3$. Contours mark
    significant values of $\alpha$. We find that $\alpha\approx 1$
    (white regions) for a large range of values, especially in the
    strong shock regime.}
  \label{fig:qFactor}
\end{figure}

\section{Cold Fronts and Shocks Generated by Colliding Streams}\label{sec:coldFronts}
In this section we present a simple analytic toy-model demonstrating
that how a collision between an inflowing stream with the ambient gas
or with another stream can lead to the formation of shocks and CFs,
and what is the resultant spatial configuration of the shocks and
CFs. Readers interested only in the simulation results may safely skip
this section.

In clusters possessing deeply penetrating streams, the inflowing
streams can collide with each other or with the ambient gas, leading
to the formation of shocks and CFs. At the moment of collision, we
find two regions whose relative velocity is converging.

As an illustrative case, consider a 1-Dimensional problem of two
compressible gaseous regions (ideal gases with a similar equation of
state), marked by $1$ and $2$, each of constant density $\rho_i$,
pressure $P_i$ and velocity $U_i$, with subscript $i$ for regions $1$
and $2$, a configuration commonly known as the Riemann Problem
\citep{Zeldovich1967}. We assume that the regions are colliding,
$\Delta U=U_2-U_1<0$, so there always exists a frame of reference in
which the velocities are of opposite sign
(\cref{fig:collision_initial}). Two of the four cases of the Riemann
problem apply for this velocity condition: either two shocks form,
propagating in opposite directions
(\cref{fig:collisionPressure_later,fig:collisionDensity_later}, or a
shock and a rarefaction wave form, also propagating in opposite
directions \citep{Zeldovich1967}. As we shall see in \rfsec{results},
the typical high inflow velocities of the streams ($\sim
1000\units{km\,s^{-1}}$) and the details of the collisions make the
double shock configuration the preferred scenario in the central
regions of clusters. In the following paragraphs we solve the
double-shock scenario in detail.

We can solve the system by treating each of the shocks separately and
utilizing the Rankine-Hugoniot shock jump conditions
\citep{Landau1959} to find the state of the post shock gas behind each
shock, which we mark as zone $3_1$ and $3_2$.

The jump conditions for the shocks propagating into zones $i={1,2}$
are
\begin{align}
P_{3,i} &= P_i\frac{2\gamma \mach_i^2-(\gamma-1)}{\gamma+1} \label{eq:rhP}\\
U_{3,i} &= U_i + \frac{P_{3,i}-P_i}{\rho_i(V_i-U_i)}\label{eq:rhU} \\
\rho_{3,i}&=\rho_i \frac{U_i-V_i}{U_{3,i}-V_i}\label{eq:rhRo} , 
\end{align}
where $V_{1,2}$ are the velocities of the two shocks, the Mach
numbers are defined as the ratio of the velocity to the sound speed
for each region $\mach_i\equiv U_i/c_i$ and the speed of sound in the
pre-shock region given by
\begin{equation}\label{eq:csound1}
c_i=\sqrt{\pderiv{P}{\rho}\Bigg|_S}=\sqrt{\gamma\frac{P_i}{\rho_i}}.
\end{equation}
The shock velocities for the two shocks are
\begin{align}
V_1&=U_1-\mach_1 c_1  \label{eq:vshock1} \\
V_2&=U_2+\mach_2 c_2     \label{eq:vshock2},
\end{align}
with the difference in sign for the second term in
\cref{eq:vshock1,eq:vshock2} due to the shocks propagating in opposite
directions.

Since there is no inherent time or length scale in this problem, the
solution for it will be self-similar, which entails that the
post-shock gas must be in pressure equilibrium, and of uniform
velocity \citep{Zeldovich1967}
\begin{align}
  P_3&=P_{3,1}=P_{3,2}  \label{eq:pCond}  \\
  U_3&=U_{3,1}=U_{3,2} \label{eq:uCond} .
\end{align}

Under these constraints, the jump conditions
\cref{eq:rhP,eq:rhU,eq:rhRo}, can be restated resulting in the
following implicit equations which can be solved numerically to obtain
the Mach numbers for the shocks
\begin{equation}\label{eq:solvP}
\frac{2\gamma \mach_2^2-(\gamma-1)}{2 \gamma  \mach_1^2-(\gamma-1)}=\frac{P_1}{P_2}
\end{equation}
\begin{multline}\label{eq:solvDU}
 \frac{\gamma-1}{2}\left(U_1-U_2\right)= -\frac{\gamma-1}{2}\Delta U=  \\
  c_1\left(\mach_1-\frac{1}{\mach_1}\right)+c_2\left(\mach_2-\frac{1}{\mach_2}\right),
   \end{multline}
where we recall that we have set $\Delta U<0$.

Once the Mach numbers are known, the post shock velocity, \cref{eq:uCond} can be solved for 
\begin{multline}\label{eq:solvU}
  U_3=\frac{U_1+U_2}{2}+ \\
    \frac{1}{\gamma-1}\left[c_1\left(\mach_1-\frac{1}{\mach_1}\right)-c_2\left(\mach_2-\frac{1}{\mach_2}\right)\right] . 
\end{multline}
In addition, we define the density contrast across the contact
discontinuity as the ratio between the two densities in the post-shock
regions as
\begin{equation}\label{eq:solvQ}
q\equiv \frac{\rho_{3,1}}{\rho_{3,2}}=\frac{\rho_1}{\rho_2} \frac{\gamma-1+2 \mach_2^{-2}}{\gamma -1+2 \mach_1^{-2}}.
\end{equation}
The density contrast is a natural measure for the prominence of the
contact discontinuity. 

We find that a contact discontinuity ($q\ne 1$), i.e.\@ a CF, should
occur whenever a stream collides, either with other streams or with
the ambient gas in the cluster. The CF will form at the Lagrangian
location of the initial interface (\cref{fig:collisionDensity_later})
and travel at the post-shock velocity $U_3$. Only in the very unlikely
case of $\rho_1=\rho_2$ and $P_1=P_2$ a CF will not form.

We now explore the relation of the density contrast of the CF, $q$ to
the initial density contrast between the streams $\rho_1/\rho_2$. We
re-write \cref{eq:solvQ} as
\begin{equation}\label{eq:solvQ1}
  q=q_0 \alpha\left(\mach_1,\mach_2;\gamma\right), 
\end{equation}
with $q_0=\rho_1/\rho_2$ and
\begin{equation}\label{eq:solvQ2}
\alpha\left(\mach_1,\mach_2;\gamma\right)= \frac{\gamma-1+2
  \mach_2^{-2}}{\gamma -1+2 \mach_1^{-2}}.
\end{equation}
It is important to note that $\mach_1$ and $\mach_2$ are \emph{not}
independent of $q_0$ and that this formulation is useful for deducing
the initial density contrast by observing the system at a developed
state.

In \cref{fig:qFactor} we plot the value of $\alpha$ for a range of
Mach numbers for the two shocks. Naturally, the high density region is
always on the side of the stronger shock. For a large range of values,
especially in the strong shock regime ($\mach>10$),
$\alpha\approx1$. The maximal and minimal values of $\alpha$ are
obtained when one shock is very strong and the other very weak
\begin{equation}\label{eq:alphaMinMax}
\frac{\gamma+1}{\gamma-1}<\alpha <\frac{\gamma-1}{\gamma+1},
\end{equation}
which for $\gamma=5/3$ is $1/4<\alpha<4$. We note that shocks found in
the central regions of the ICM are typically weak shocks.

The solution presented above is valid for most reasonable collision
situations, however it is possible to construct initial conditions
where this solution is not valid. In the limit of $\Delta U\to0$,
\cref{eq:solvDU} can only be satisfied for $\mach_1 \approx \mach_2
\approx1$, a condition that cannot be satisfied simultaneously with
\cref{eq:solvP} for all values of $P_1/P_2$. The physical
interpretation of this is that, assuming $P_2>P_1$, for a given value
of $P_1/P_2$, there exists a critical velocity difference $\Delta
U_C$, such that for collisions with lower velocity difference (in
absolute value) the resulting post-shock pressure is $P_3<P_2$.

Under these conditions, the second case of the Riemann problem becomes
relevant, namely a single shock wave propagating into zone $1$ (since
we assumed $P_1<P_2$) which raises the pressure to $P_3$ in the
post-shock gas, and a rarefaction wave propagating into zone $2$ which
\emph{lowers} the pressure to $P_3$, as in the solution to the `shock
tube problem' (see e.g.\@ \citealt{Zeldovich1967}, Chap.\@ 4). The
medium between the shock wave and rarefaction wave is also in pressure
equilibrium and contains a contact discontinuity, but its density
contrast will naturally \emph{not} be given by \cref{eq:solvQ}. As we
shall see, this scenario is not expected in collision sites of the
streams in the central regions of the clusters, due to the very high
velocities associated with the streams.

The 1D scenario explored in this section is very idealized and does
not address many of the properties and processes found in the ICM such
as turbulent motions, radiative cooling, gravity etc.\@, but as we
shall see, it still captures the essential aspects of the process
leading to the formation of CFs in the ICM.

\begin{table}
\centering
  	 \begin{tabular}{@{}lcccccc@{}}
	 \toprule
	 Cluster & Redshift &$\Mv$&$\Rv$&$\Tv$&$\Vv$\\ 
	  & z &$[10^{14}\msun]$&$[\unitstx{Mpc}]$&$[10^7\units{K}]$&$[\unitstx{km\, s^{-1}}]$ \\ 
        \midrule
	
  	CL6 & 0  & 3.3  & 1.80 & 2.9 & 894 \\
        CL6 & 0.6   & 2.4  & 1.15 & 3.2 & 946 \\
        CL107 &  0 & 6.6  & 2.26 & 4.5  & 1125 \\
 	\bottomrule 
	\end{tabular}
	\caption{Properties of the dynamically active clusters CL6 at \zeq{0} and
          \zeq{0.6} and CL107 at \zeq{0}. Virial quantities were calculated for an
          over-density of $\delvir=337$.}
	 \label{tab:clusterProperties}
 \end{table}

\section{Simulations}\label{sec:sims}
The simulations were carried out with the Adaptive Refinement Tree
(\textsc{art}) \nbody+gas-dynamics code \citep{Kravtsov1999}, an
Eulerian code that uses adaptive refinement in space and time, and
(non-adaptive) refinement in mass \citep{Klypin2001} to reach the high
dynamic range required to resolve cores of haloes formed in
self-consistent cosmological simulations.

The systems were extracted from cosmological simulations in a flat
$\Lambda$CDM model: $\Omega_{\mathrm{m}}=1-\Omega_{\Lambda}=0.3$,
$\Omega_{\mathrm{b}}=0.04286$, $h=0.7$, and $\sigma_8=0.9$, where the
Hubble constant is defined as $100h\units{km\,s^{-1}\,Mpc^{-1}}$, and
$\sigma_8$ is the power spectrum normalization on an
$8h^{-1}\units{Mpc}$ scale.  The computational boxes of the
large-scale cosmological simulations were either $80
h^{-1}\units{Mpc}$ (CL6, see \rfsec{cl6}) or $120 h^{-1}\units{Mpc}$
(CL107, see \rfsec{cl107}), and the simulation grid was adaptively
refined to achieve a peak spatial resolution of $\sim 7h^{-1}$ and
$5h^{-1}\units{kpc}$ respectively. These simulations are described in
detail in \citet{Kravtsov2006}, \citet{Nagai2007} and
\citet{Nagai2007a}. Adaptive mesh refinement techniques, such as the
one employed in the simulation, are particularly suited to capture
discontinuous features such as shocks and contact discontinuities
which make it especially suitable for our purposes.

The simulation suite analysed in this study is comprised of $16$
cluster-sized systems at \zeq{0} spanning a mass range of $ 8.6 \times
10^{13} \textrm{--} 2.2 \times 10^{15} \msun$, and their most massive
progenitors at \zeq{0.6}. In this paper we have chosen to focus on two
clusters, CL6, in which a clear and compelling example of our proposed
mechanism is realized (\rfsec{cl6}), and CL107 in which the compound
effect of a stream collision and a satellite give rise to shocks and CFs
(\rfsec{cl107}).

The gas in the simulations is treated as a mono-atomic ideal gas, and
is thus well described by an equation of state
\begin{equation}\label{eq:eos}
 P=(\gamma-1)\rho e,
\end{equation}
where $P$, $\rho$ and $e$ are the pressure, density and specific
internal energy respectively and $\gamma$ is the adiabatic index
($\gamma=5/3$ for a mono-atomic gas).

Besides the basic dynamical processes of collision-less matter (dark
matter and stars) and gas-dynamics, several physical processes
critical for galaxy formation are incorporated: star formation, metal
enrichment and feedback due to Type II and Type Ia supernov\ae, and
self-consistent advection of metals. The cooling and heating rates
take into account Compton heating and cooling of plasma, heating by
the UV background \citep{Haardt1996}, and atomic and molecular
cooling, which is tabulated for the temperature range $10^2$ to $10^9
\units{K}$, a grid of metallicities, and UV intensities using the
\textsc{cloudy} code (version 96b4; \citealt{Ferland1998}). The
\textsc{cloudy} cooling and heating rates take into account
metallicity of the gas, which is calculated self-consistently in the
simulation, so that the local cooling rates depend on the local
metallicity of the gas. The star formation recipe incorporated in
these simulations is observationally motivated (e.g.\@
\citealt{Kennicutt1998}) and the code also accounts for the stellar
feedback on the surrounding gas, including injection of energy and
heavy elements (metals) via stellar winds, supernov\ae, and secular
mass loss.

The simulations do \emph{not} include AGN feedback mechanisms, and
while this may lead to unrealistic conditions in the core of the
cluster, it allows us to isolate the role of the gas streams in
determining the conditions in the inner regions of the ICM from that
of the AGN.

The virial quantities of the mass, radius, temperature and velocity
($\Mv,\Rv,\Tv,\,\&\,\Vv$) of the clusters are defined for an
over-density $\delvir=337$ at \zeq{0} and $\delvir=224$ at \zeq{0.6}
\citep{Bryan1998} above the mean density of the universe. The
properties of the dynamically active `unrelaxed' clusters at \zeq{0}
and \zeq{0.6} are summarized in \cref{tab:clusterProperties}.

The simulations of the clusters were classified visually as `relaxed' or
`unrelaxed'. The classification was carried out to emulate as close as
possible the methods employed by observational studies. The classification is
described in detail in \citet{Nagai2007a,Nagai2007}, and is based on mock
\emph{Chandra} X-ray images of the clusters.  Based on the mock observations,
clusters were classified as relaxed if they possessed regular X-ray morphology
and a single luminosity peak, with minimal deviation of the isophotes from
elliptical symmetry. In contrast, unrelaxed clusters are those with secondary
luminosity peaks, filamentary X-ray structures, or significant shifts in the
centres of the isophotes. A cluster was deemed unrelaxed if it appeared so in
at least one of the 3 orthogonal Cartesian projections. Within our simulation
suite, 10 clusters are identified as unrelaxed and 6 as relaxed at \zeq{0}
\citep[see Table 1 of][]{Nagai2007}. It is important to note that we find
deeply penetrating streams only in unrelaxed clusters \citep[see Fig.\@ 12
of][]{Zinger2016}.

In this work, we focus on analysing the dynamically
active clusters that exhibit prominent features of shocks and CFs.

The gas velocities are shown with respect to an inertial frame defined
by a centre-of-mass velocity of the cluster. For each cluster in the
suite, a centre-of-mass velocity profile was calculated for the 3
velocity components and a radius was selected where the profiles were
seen to level off, typically at $\sim 1-2 \Rv$, and remain nearly
constant beyond that. The gas velocities are then measured with
respect to the centre-of-mass velocity as calculated for the gas
within that radial limit. As a result, the centre-of-mass velocity is
largely insensitive to the choice of the radial limit.

\begin{figure}
  \centering
  \subfloat[Gas Density]{\label{fig:cl6Map8_gas}
    \includegraphics[width=8.56 cm,keepaspectratio,bb=0 0 5.33in 4.43in]{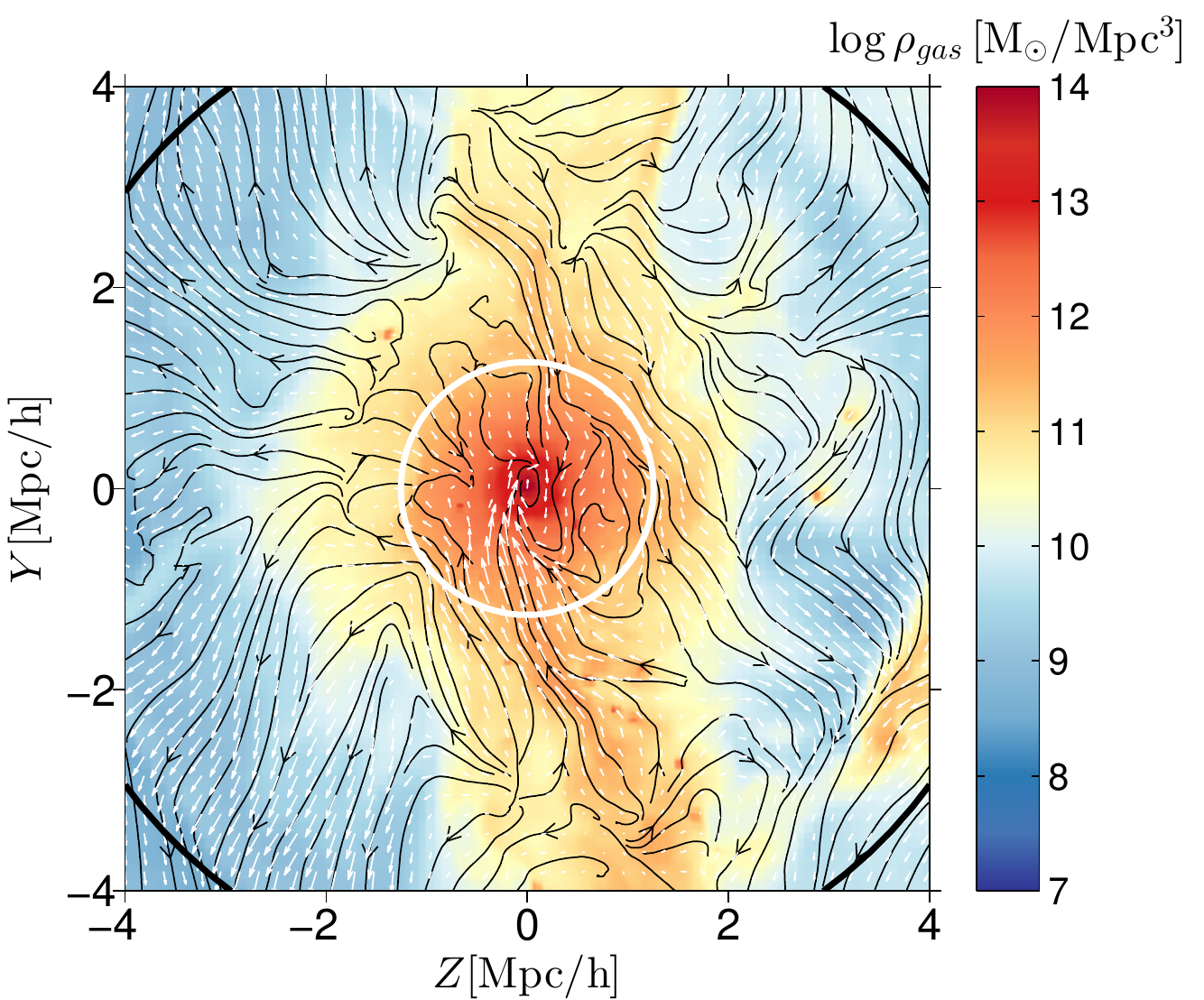}}\\
  \subfloat[Dark matter Density]{\label{fig:cl6Map8_dm}
    \includegraphics[width=9 cm ,keepaspectratio,bb=0 0 5.36in 4.43in]{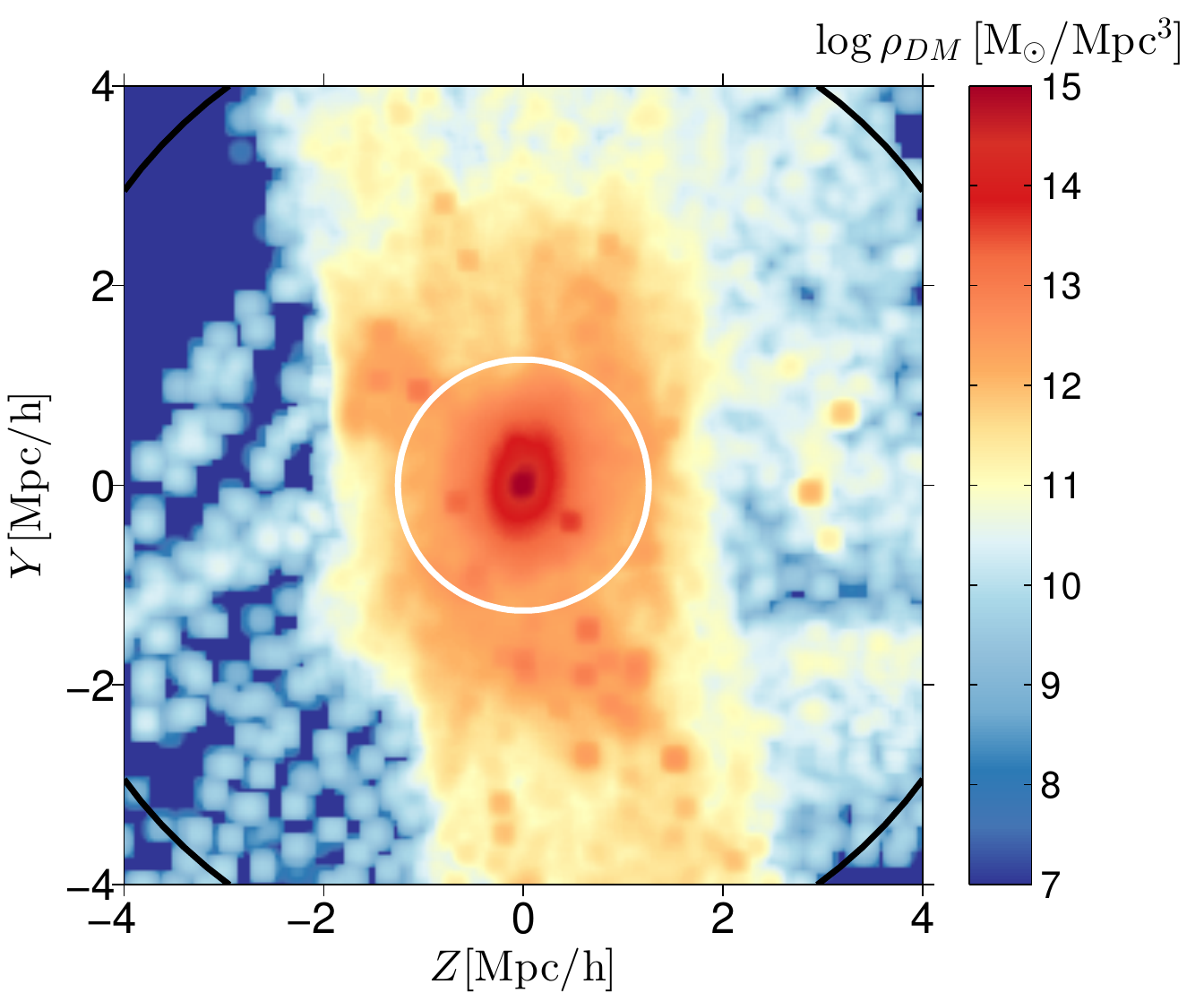}}
  \caption{Gas and dark matter density of the cluster CL6 at \zeq{0},
    \subrf{fig:cl6Map8_gas} \& \subrf{fig:cl6Map8_dm}, respectively.
    The white circles denote the virial radius of the cluster
    and the black circles demark the approximation location of
    the virial accretion shock (in this projection). The velocity
    field is shown as arrows (white) and streamlines
    (black) in \subrf{fig:cl6Map8_gas}. The values shown are
    averaged over a slice of $ 500h^{-1} \units{kpc}$. }
  \label{fig:cl6Map8}
\end{figure}

\section{Colliding Streams in Simulated Clusters}\label{sec:results}
In this section we present a representative example of two simulated
clusters possessing deeply penetrating streams in which the
collision between inflowing streams can be seen to create shocks and
CFs in the ICM.

As we have seen, the CF interface is in pressure equilibrium. Since
for an ideal gas equation of state $P\propto \rho T$, this entails
that the temperature and density contrasts are inversely proportional
to each other:
\begin{equation}\label{eq:cfRhoT}
P_1=P_2\to\, \rho_1T_1=\rho_2T_2 \to\, \frac{T_1}{T_2}=\left(\frac{\rho_1}{\rho_2}\right)^{-1},
\end{equation}
where the subscripts $1$ and $2$ denote the conditions on either side
of the interface.  We use the proxy for entropy $S\propto
T\rho^{-2/3}$ and find that it is an excellent indicator of CFs in
simulations
\begin{equation}\label{eq:cfEntropy}
\frac{S_1}{S_2}=
\frac{T_1}{T_2}\left(\frac{\rho_1}{\rho_2}\right)^{-\frac{2}{3}}=
\left(\frac{T_1}{T_2}\right)^{\frac{5}{3}}=
\left(\frac{\rho_1}{\rho_2}\right)^{-\frac{5}{3}}=q^{-\frac{5}{3}} .
\end{equation}

\subsection{Colliding Streams in CL6}\label{sec:cl6} 

The cluster CL6 is a $\Mv=3.3\times 10^{14}\units{\msun}$ cluster with
a virial radius of $\Rv=1.8\units{Mpc}$ at \zeq{0}. \cref{fig:cl6Map8}
shows the gas and dark matter density of the cluster on a scale of
several $\Rv$. The cluster is situated along a large scale dark matter
filament, whose diameter is $\sim \Rv$ \citep{Dekel2006}. The
accretion shock of the ICM extends to $\sim 4 \Rv$
\citep{Lau2015,Zinger2016b}. The gas accretion can be seen to occur
predominantly through streams which lie along the centre of the dark
matter filament.

\begin{figure}
  \centering
  \includegraphics[width=9cm,keepaspectratio,bb=0 0 5.07in 4.43in]{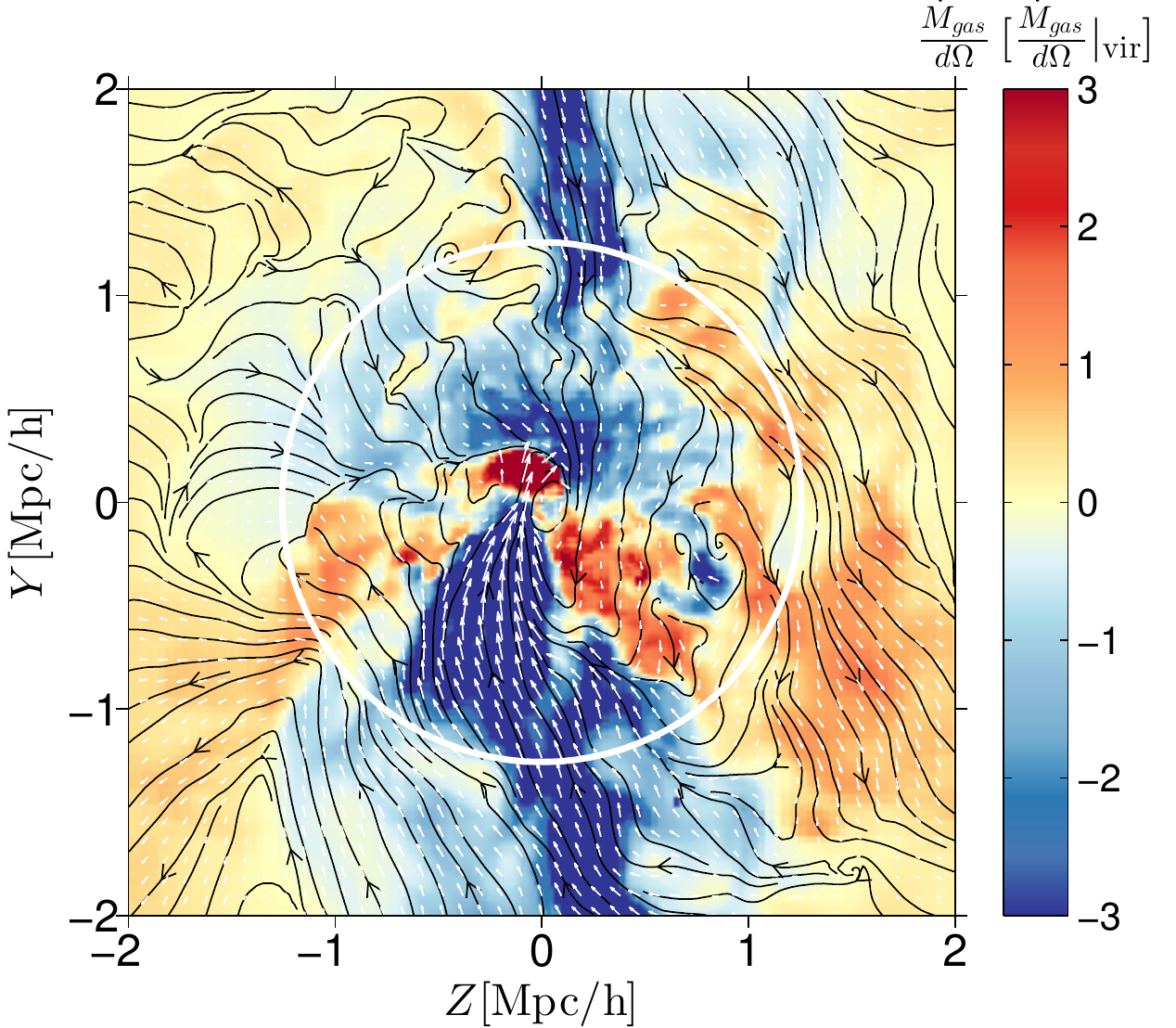}
  \caption{The gas mass accretion rate into the simulated cluster
    CL6 at \zeq{0}. The virial radius is shown as a white circle. The
    velocity field is shown as arrows (white) and streamlines
    (black). The values shown are averaged over a slice of
    $ 100h^{-1} \units{kpc}$. A gas stream is flowing from the bottom
    upwards, penetrating into the centre and colliding with the
    ambient gas. A second stream flowing from the top stops just
    inside the virial radius.}
  \label{fig:cl6Flux}
\end{figure}

In \cref{fig:cl6Flux} we show the gas mass accretion rate, defined as
\begin{equation}\label{eq:massflux} 
\dot M(r)=\int_{\Omega} \rho \left[\left(\vec v-\vec{v}_{cm}\right) \cdot \hat
  r\right] r^2 \diff \Omega ,
\end{equation}
for the cluster. A prominent stream enters from the bottom and flows
to the centre, where it collides with the ambient gas which is also
flowing inwards. The stream from the bottom actually overshoots the
centre (and thus the radial mass accretion rate changes sign abruptly)
before colliding with the inflowing gas coming from the top. Another
stream can be identified flowing from the top and dissipating just
inside the virial radius.

\begin{figure}
  \centering
  \includegraphics[height=8cm,keepaspectratio,bb=0 0 5.24in 4.43in]{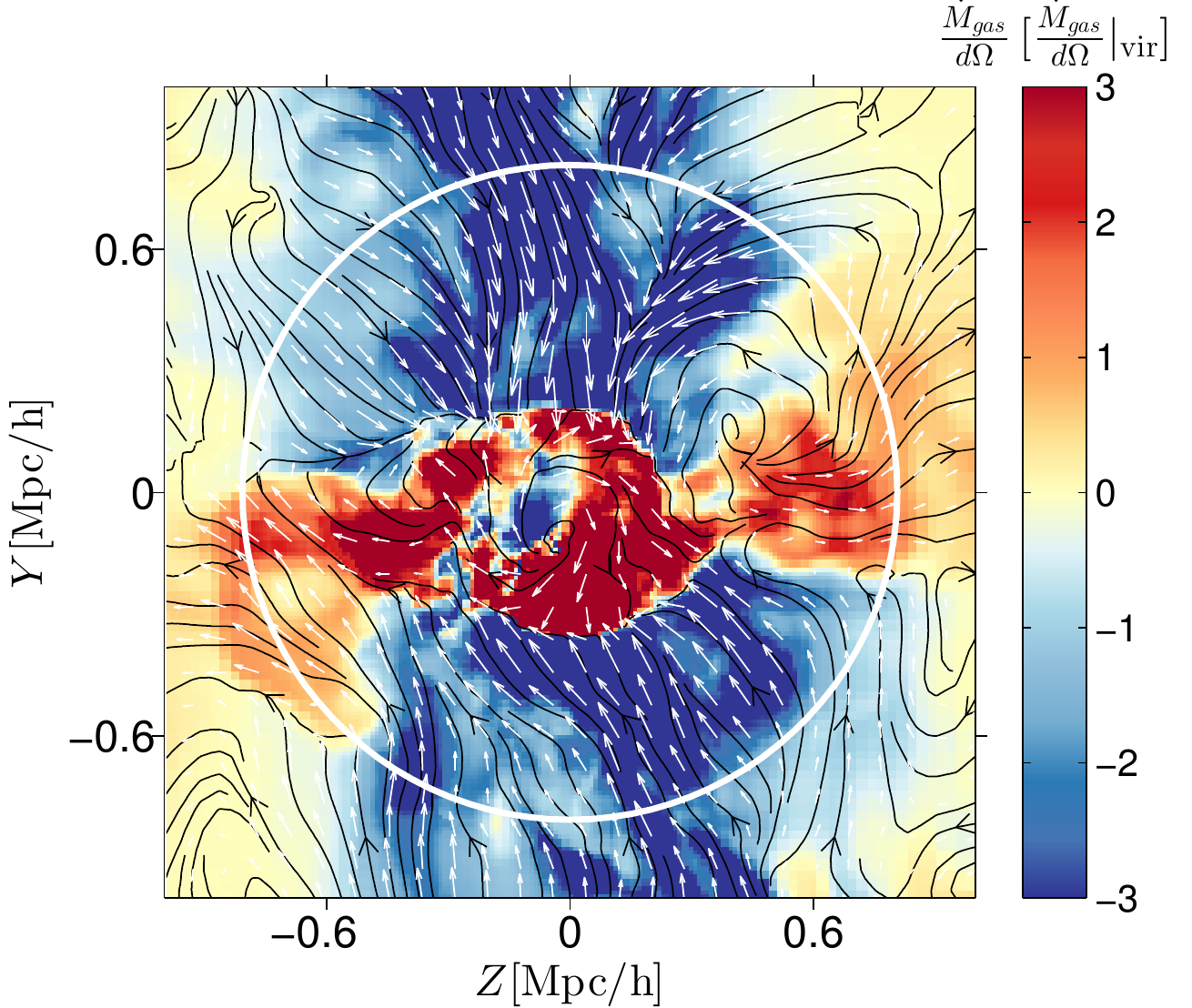}
  \caption{Mass inflow rate of CL6 at \zeq{0.6}. $\Rv$ is marked by a
    white circle, with the velocity field shown as arrows (white) and
    streamlines (black). Two streams flow into the cluster from the
    top and bottom and reach the centre where they are stopped by
    shock waves. Values are averaged over a slice of $125h^{-1}
    \units{kpc}$.}
  \label{fig:cl6Map06_flux}
\end{figure}

In \citet{Poole2006}, it was shown that during major mergers in
clusters, stripped gas can stream towards the centre in high-velocity
flows, which persist for as much as $2\unitstx{Gyrs}$ after
forming. However, we find that the streams we find in this cluster do
not originate with a merger event. The last major merger in the system
occurred in the period \zeq{1.0} to \zeq{0.5} \citep{Nagai2003}, and
any residual streams would have dissipated in the intervening
$5\units{Gyrs}$. In addition, residual streams from mergers need not
be aligned with the large-scale filaments of the cosmic web, whereas
the streams seen here and studied in depth in \citet{Zinger2016} are
large scale features, several $\unitstx{Mpc}$ in length and in some
cases as wide as the virial radius of the cluster (e.g.\@
\cref{fig:cl6Map06_flux}), which can be traced form the filaments of
the cosmic web as they penetrate into the cluster.

That having been said, since the accretion occurs predominately along
the gas streams, most of the merging substructure also flows along the
streams resulting in a complex interplay between smooth and clumpy
accretion.

\begin{figure}
  \centering
  \subfloat[CL6 Velocity]{\label{fig:cl6Vel_b2}
    \includegraphics[width=9cm,keepaspectratio,bb=0 0 5.35in 4.42in]{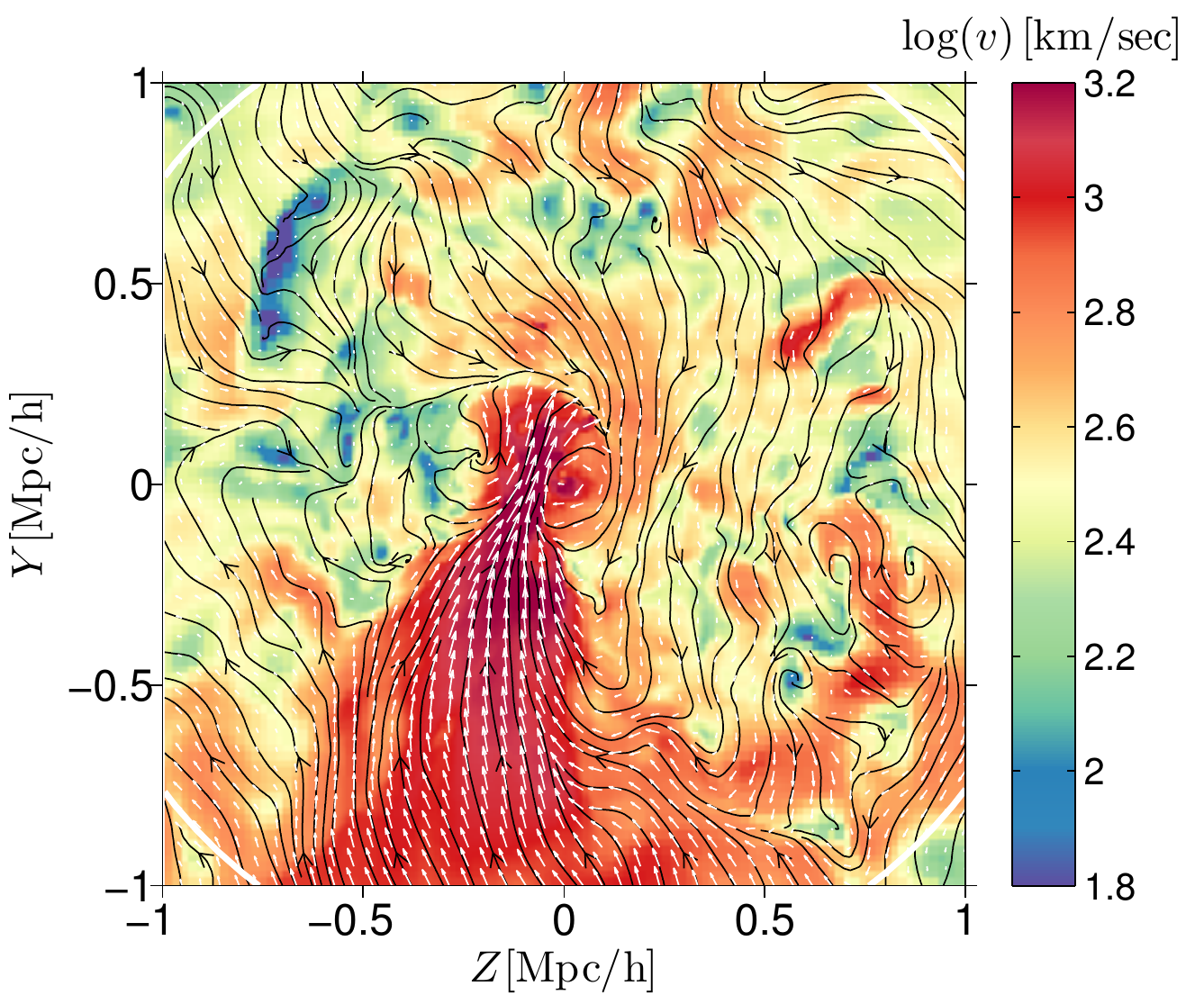}}\\
  \subfloat[CL6 Mach Number]{\label{fig:cl6Mach_b2}
    \includegraphics[width=8.57cm,keepaspectratio,bb=0 0 5.1in 4.36in]{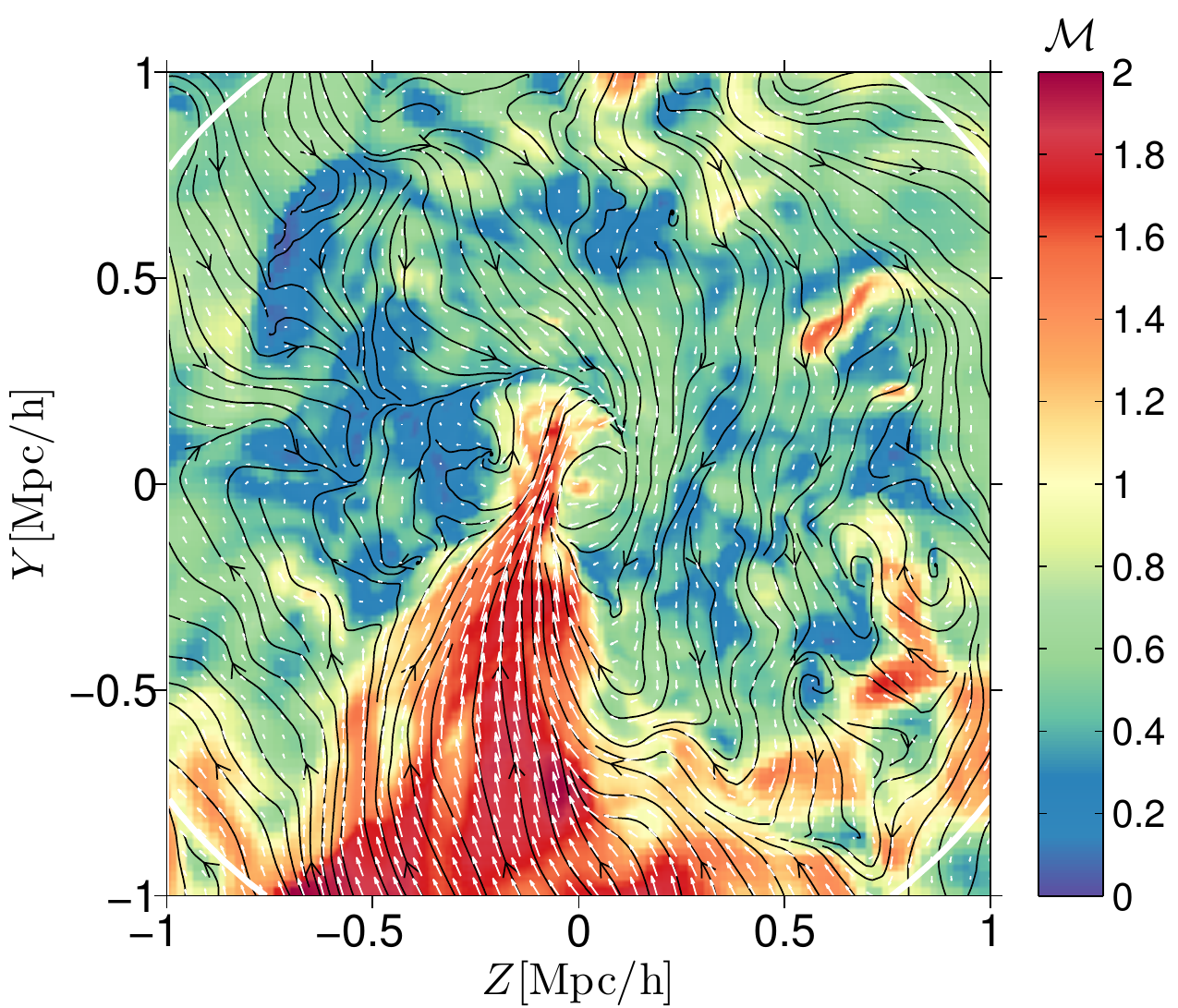}}
  \caption{The magnitude of gas velocity in the simulated cluster CL6
    at \zeq{0} is shown in \subrfig{cl6Vel_b2} and the corresponding
    Mach number $\mach=v/c_{\mathrm{s}}$ for the flow shown in
    \subrfig{cl6Mach_b2}. The virial radius is shown as a white circle
    (seen at the corners of the plots). The velocity field is shown as
    arrows (white) and streamlines (black). The values shown are
    averaged over a slice along the $X$ plane of $ 50h^{-1}
    \units{kpc}$. A high velocity \mbox{($\gtrsim
      1000\units{km\,s^{-1}}$)} supersonic stream is flowing upwards,
    overshooting the centre and colliding with the ambient gas which
    is flowing inwards.}
  \label{fig:cl6VelMach}
\end{figure}

In \cref{fig:cl6VelMach} we show the local velocity of the gas and its
corresponding Mach number $\mach=v/c_{\mathrm{s}}$, where the typical sound speed
is defined for a shell of given radius as
\begin{equation}\label{eq:csonic}
c_{\mathrm{s}}(r) = \sqrt{\pderiv{P}{\rho}\Bigg|_S} =\sqrt{\gamma
  \frac{\kboltz\overline{T}}{\mu m_{\mathrm{p}}}} .
\end{equation}
The second expression derives from the ideal gas equation of state
used to describe the gas. $\kboltz$ is the Boltzmann constant, $\mu
m_{\mathrm{p}}\simeq 0.59 m_{\mathrm{p}}$ is the average particle mass ($m_{\mathrm{p}}$ is the proton
mass) and $\overline{T}$ is the mass-weighted mean temperature of a
spherical shell of radius $r$.

The typical velocities are of order $500\units{km\,s^{-1}}$
for the gas coming from the top and $1000 \units{km\,s^{-1}}$ for the bottom
stream, which is moving supersonically with respect to its
surrounding. The velocity difference in the collision is of order
$\sim 1500\units{km\,s^{-1}}$. The Mach numbers of the collisions are thus $1.5-2$.
\begin{figure*}
  \centering
  \subfloat[Velocity]{\label{fig:cl6Maps_vel}
    \includegraphics[height=6.3cm,keepaspectratio,bb=0 0 5.35in 4.42in,trim=0.70in 0.42in -0.06in  0.06in, clip]{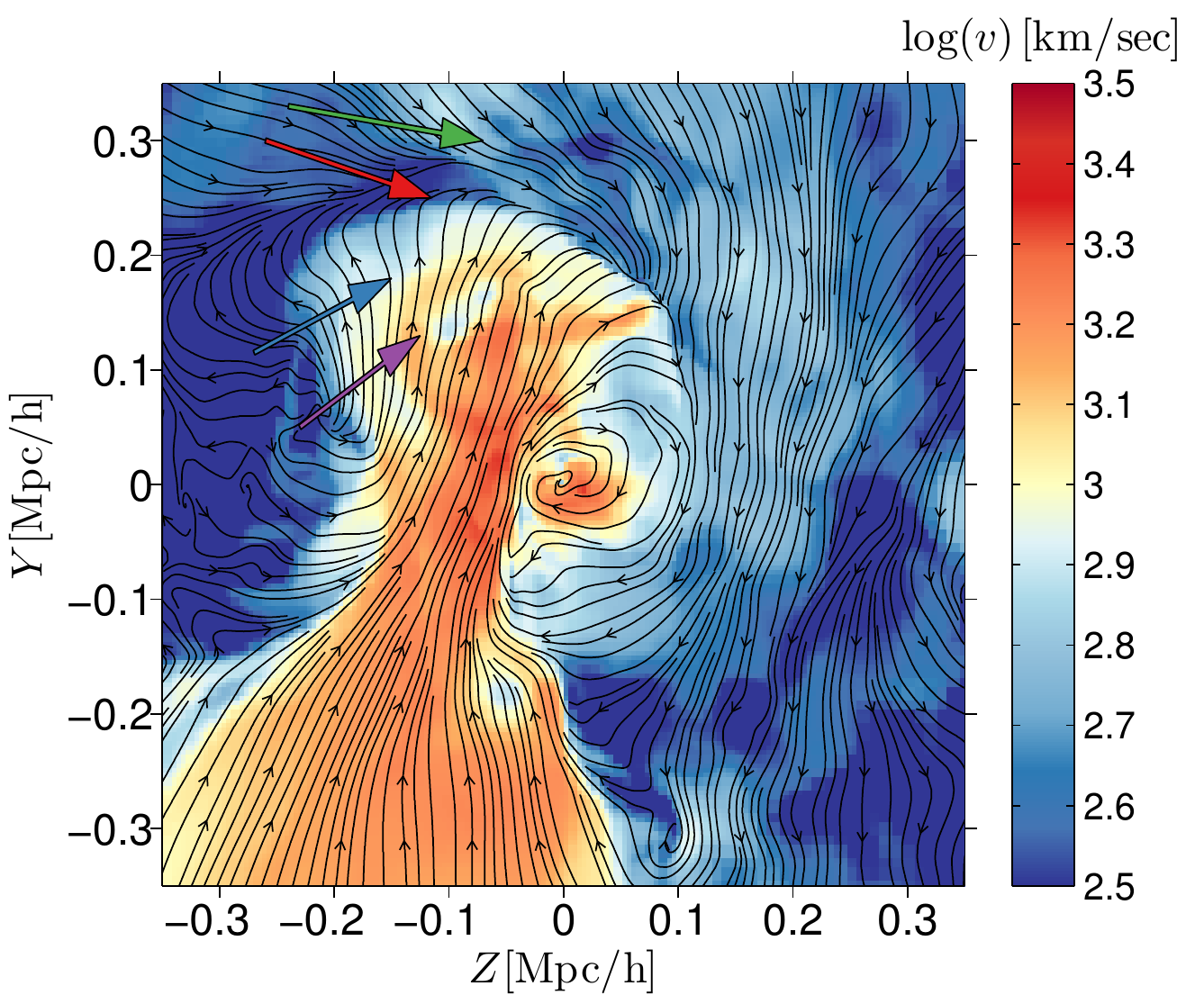}}
  \subfloat[Metallicity]{\label{fig:cl6Maps_zmet}
    \includegraphics[height=6.3cm,keepaspectratio,bb=0 0 5.19in 4.42in,trim=0.7in 0.42in -0.22in  0.06in, clip]{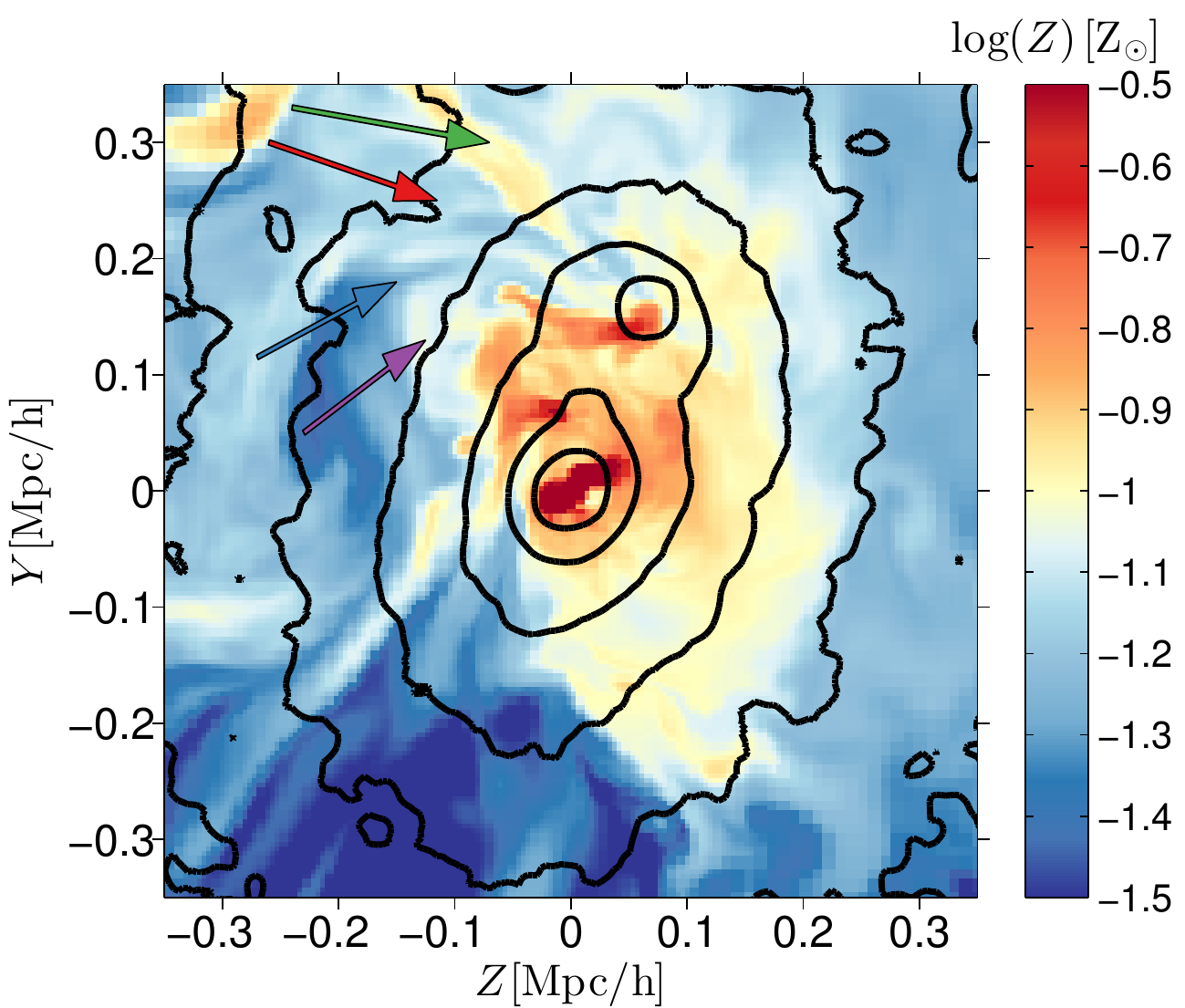}}\\
  \subfloat[Entropy]{\label{fig:cl6Maps_ent}
    \includegraphics[height=6.3cm,keepaspectratio,bb=0 0 5.36in 4.43in,trim=0.7in 0.42in -0.05in  0.06in, clip]{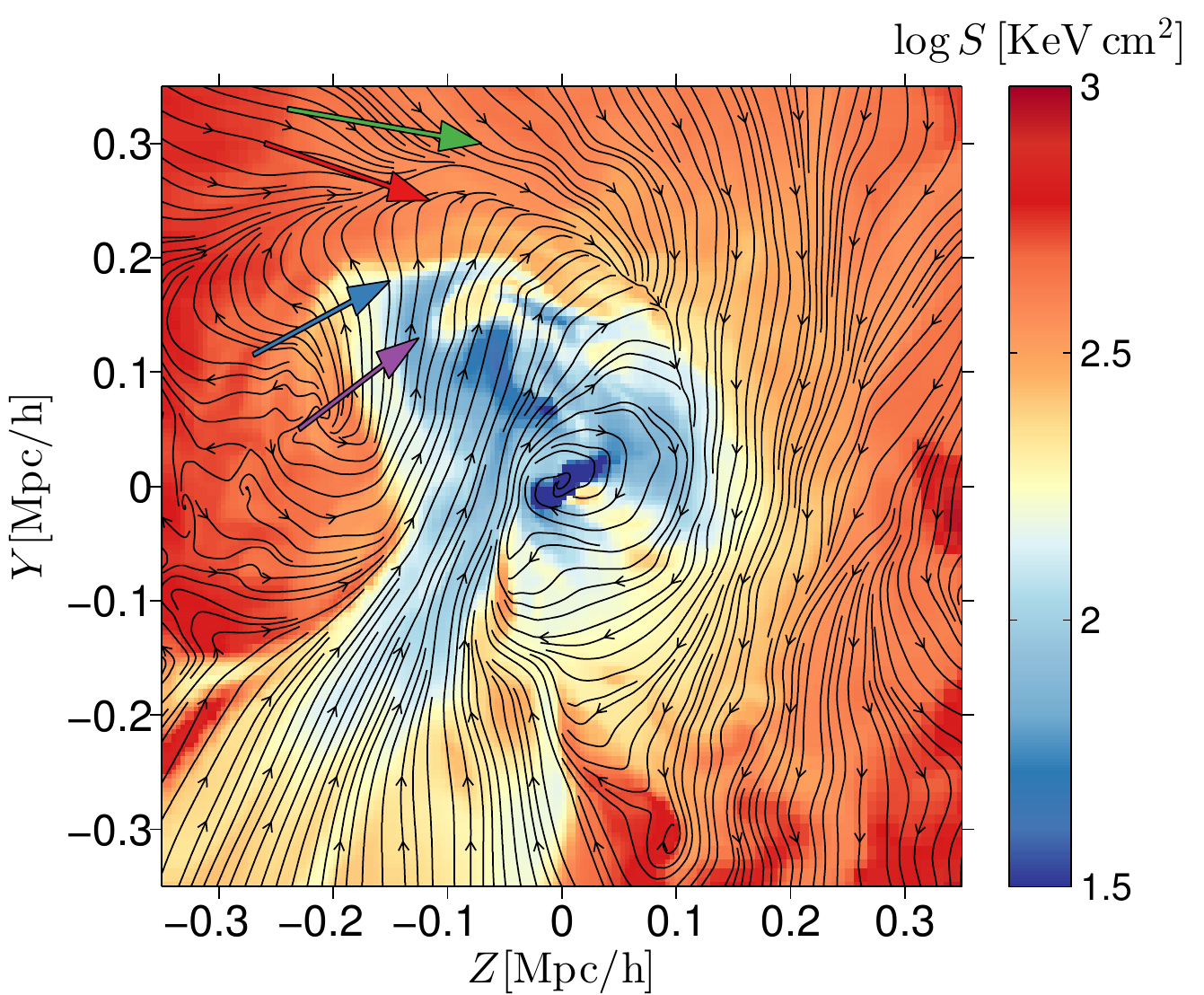}}
  \subfloat[Temperature]{\label{fig:cl6Maps_temp}
    \includegraphics[height=6.3cm,keepaspectratio,bb=0 0 5.1in 4.42in,trim=0.7in 0.42in -0.31in  0.06in, clip]{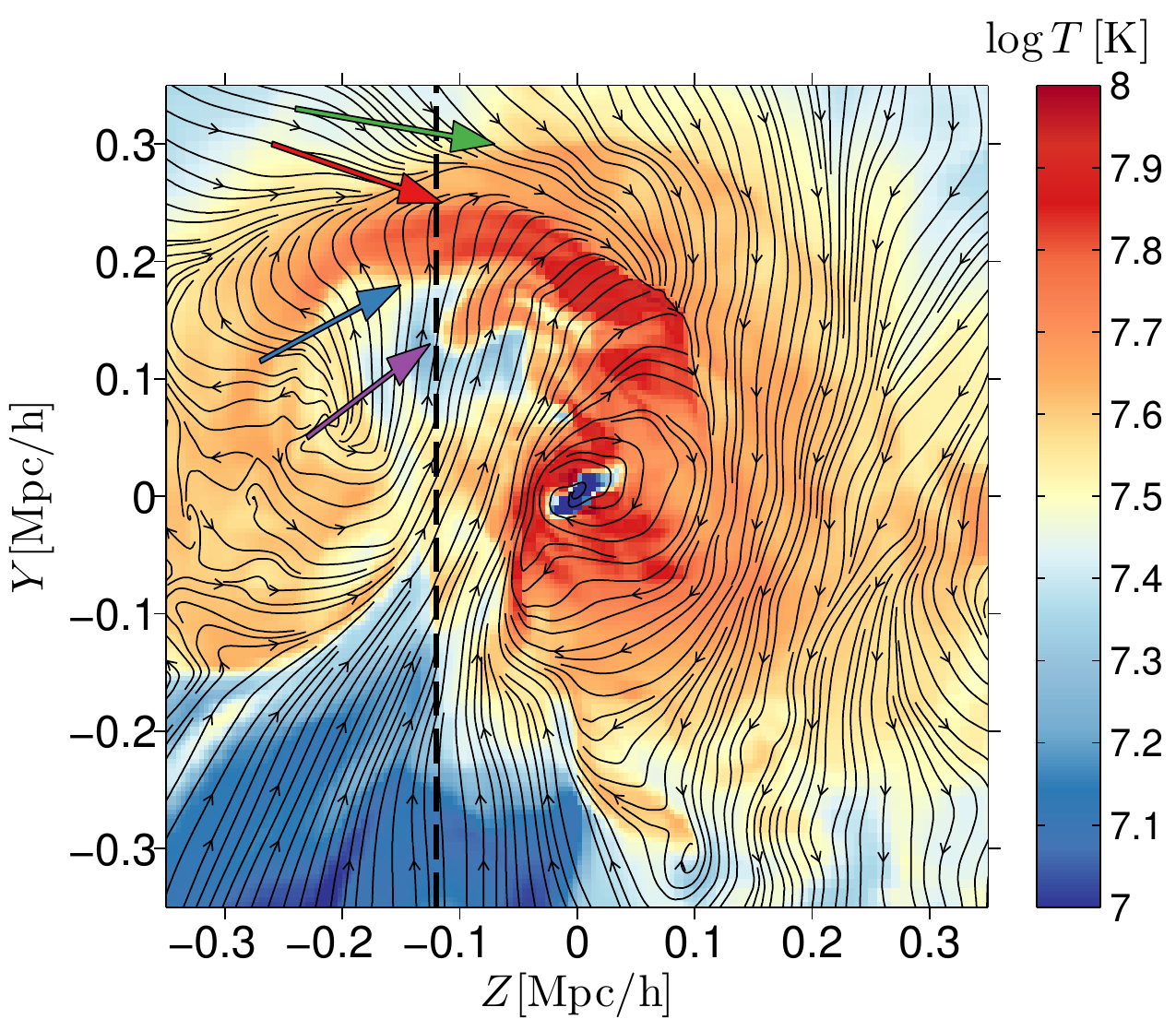}}\\
  \subfloat[Pressure]{\label{fig:cl6Maps_press}
    \includegraphics[height=6.3cm,keepaspectratio,bb=0 0 5.19in 4.42in,trim=0.7in 0.42in -0.22in  0.06in, clip]{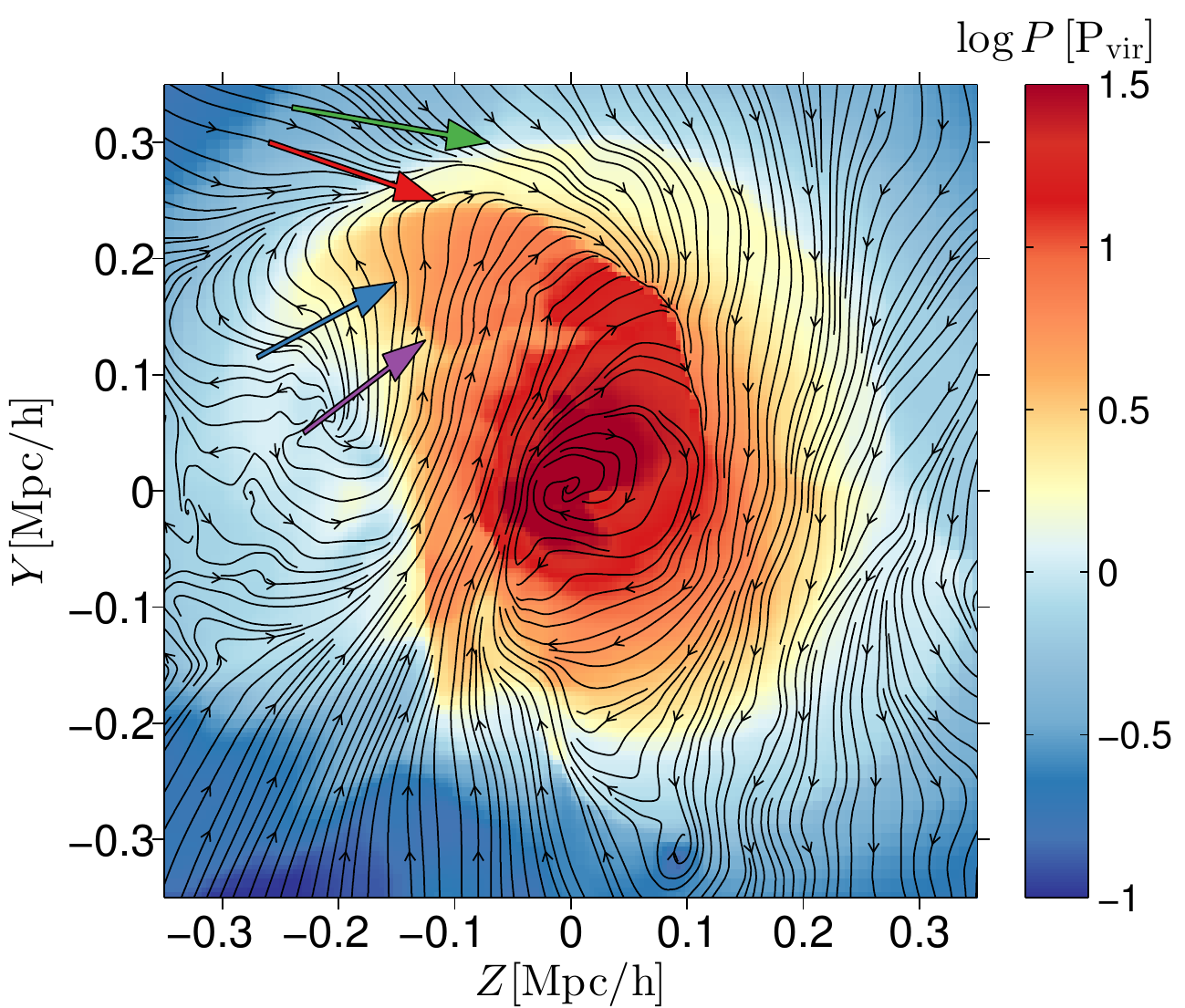}}
  \subfloat[Gas Density]{\label{fig:cl6Maps_rho}
    \includegraphics[height=6.3cm,keepaspectratio,bb=0 0 5.5in 4.43in,trim=0.7in 0.42in -0.09in  0.06in, clip]{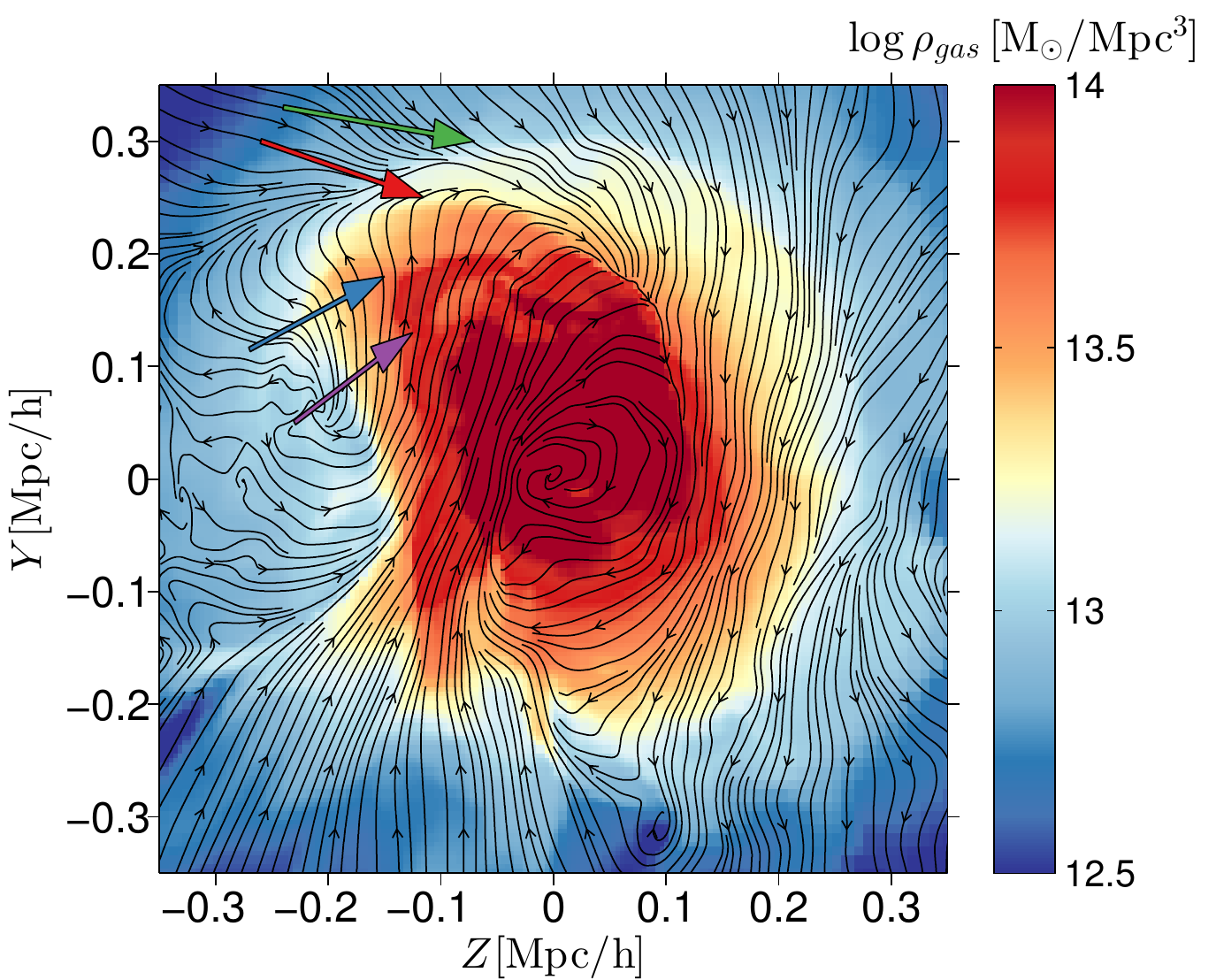}}
  \caption{The collision zone in CL6 at \zeq{0} is shown in detail in
    a box of size $0.7h^{-1} \units{Mpc}$ centred on the cluster
    centre. The velocity \subrfig{cl6Maps_vel}, metallicity
    \subrfig{cl6Maps_zmet}, entropy \subrfig{cl6Maps_ent}, temperature
    \subrfig{cl6Maps_temp}, pressure \subrfig{cl6Maps_press} and gas
    density \subrfig{cl6Maps_rho} are all averaged over a slice of
    $25h^{-1} \units{kpc}$. Dark matter density contours (black) are
    shown in \subrfig{cl6Maps_zmet} (averaged over a slice of width
    $50h^{-1} \units{kpc}$). Streamlines represent the velocity
    field. Areas of interest are marked by colored arrows as follows:
    the red arrow marks the prominent shock propagating away from the
    collision site and the green arrow marks another shock, which
    formed earlier. The blue arrow marks a CF and the purple arrow
    marks the location of another weaker shock propagating downwards
    (see \cref{fig:cl6Bore}). The dashed vertical line in
    \subrfig{cl6Maps_temp} marks the cross-section used to construct
    the profiles of the gas properties in \cref{fig:cl6Bore}.}
  \label{fig:cl6Maps}
\end{figure*}

In \cref{fig:cl6Maps} we examine in detail the area in which the
stream collides with the ambient gas, by showing the various
properties of the gas. A shock front propagating upwards is clearly
visible in the interface between the two streams (e.g.\@
\cref{fig:cl6Maps_vel}) marked by a red arrow. It appears as a sharp
discontinuity in the temperature, pressure and density maps
(\cref{fig:cl6Maps_temp,fig:cl6Maps_press,fig:cl6Maps_rho}) as well as
in the velocity field, but only barely distinguishable in the entropy
map \cref{fig:cl6Maps_ent}.

Just above this shock, another less distinct shock may be identified,
especially in the pressure and temperature maps
(\cref{fig:cl6Maps_temp,fig:cl6Maps_press}). A green arrow marks its
position. Since this shock is also propagating upwards, it stands to
reason that this shock was formed earlier, perhaps due to a change in
the penetration or direction of the gas flow.

\begin{figure}
  \centering
  \subfloat[Thermodynamic Properties]{\label{fig:cl6Bore_prof}
    \includegraphics[width=8.5cm,keepaspectratio,bb=0 0 5.18in 4.04in]{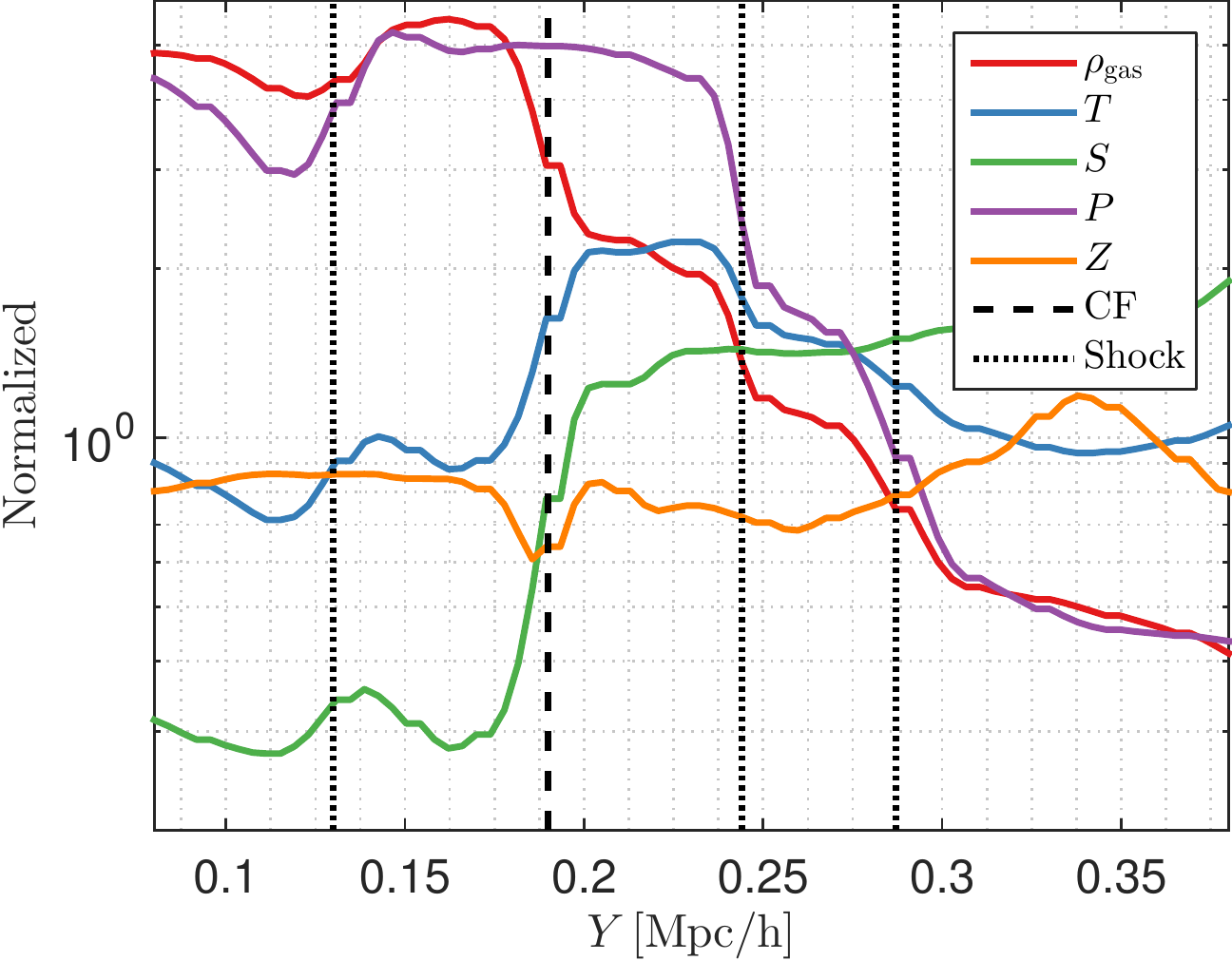}}\\
  \subfloat[Velocity Components]{\label{fig:cl6Bore_vprof}
    \includegraphics[width=8.5cm,keepaspectratio,bb=0 0 5.19in 4.15in]{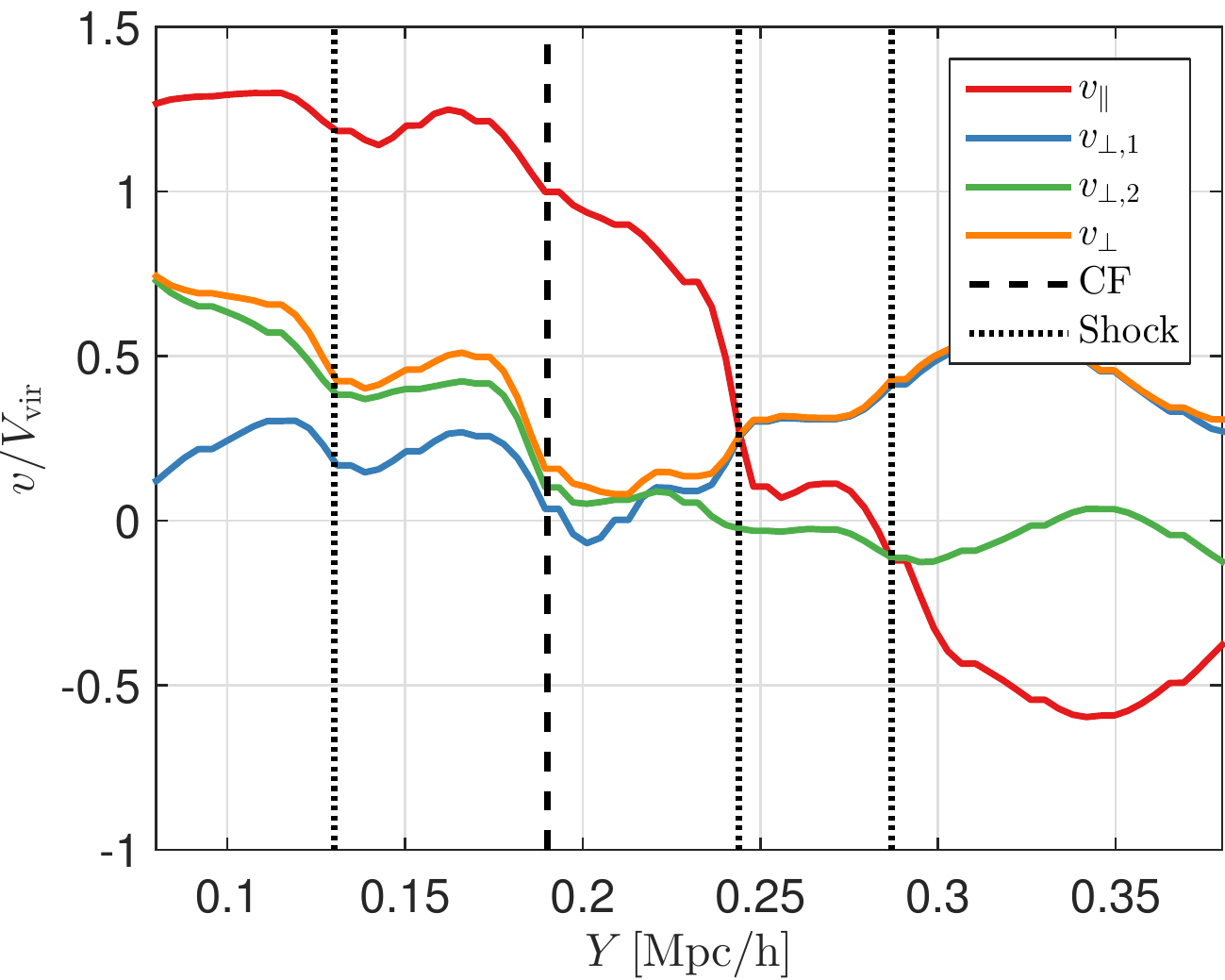}}
  \caption{Profiles of the gas properties in CL6 at \zeq{0} along a
    line in the $Y$ direction (and $X=0$), perpendicular to the shock
    and CF, at \mbox{$Z=-0.12h^{-1} \units{Mpc}$} (see
    \cref{fig:cl6Maps_temp}). Gas density (red), temperature (blue),
    entropy (green) pressure (purple) and metallicity (orange) are
    shown in \subrfig{cl6Bore_prof}. The parameters are all in units
    of the virial parameters except for metallicity, which is units of
    $0.1 Z_\odot $. The velocity components parallel (red) and
    perpendicular (orange) to the line are shown in
    \subrfig{cl6Bore_vprof}. $v_{\perp}$ is decomposed into two
    components $v_{\perp,1}=v_z$ (blue) in the plane of
    \cref{fig:cl6Maps} and $v_{\perp,2}=v_x$ (green) perpendicular to
    it. Locations of shock fronts and CFs are also marked by black
    dotted and black dashed lines, respectively. The double shock
    configuration, with CF between them, as described in
    \rfsec{coldFronts} appears to have been realized in the simulated
    cluster.}
  \label{fig:cl6Bore}
\end{figure}

Just below the shock front, a prominent CF can be identified, when
looking from the shock front downwards, as a drop in entropy and
temperature and a rise in gas density (marked by a blue arrow). As
expected, the pressure remains constant across the CF. Another feature
we suspect of being a shock is barely discernible in these maps, but
can be identified when following the profiles of the gas properties as
done in \cref{fig:cl6Bore}. Its position is marked by a purple arrow.

In \cref{fig:cl6Maps_zmet} the gas metallicity is shown as a color
map, overlaid with the dark matter density distribution, in black
contours. Just above the central density peak of the cluster a smaller
density peak can be identified. This is a sub-halo in the cluster
which hosts a satellite galaxy. A tail of high metallicity gas,
stripped by ram-pressure and tidal forces, can be seen trailing the
satellite, indicating that the satellite is moving towards the
  right, as can also be seen in the velocity field in the region.

The over-shooting stream seen here is similar to structures which
arise due to infalling satellites which can leave wakes of gas behind
them \citep[e.g.\@][]{Poole2006}, though this stream clearly
originates with the inflowing matter coming from the filaments of the
cosmic web. It is possible that a satellite, travelling along the
stream was instrumental in enabling the deep penetration of the
stream. However, the CFs seen in \cref{fig:cl6Maps} appear to be the
result of the stream collision with the ambient gas. We revisit the
connection between stream and satellites as generators of CF formation
in \rfsec{discuss}.


We study the shock fronts and CFs in detail by following the gas
properties along a vertical line in the $Y$ direction of the cluster
at a position of \mbox{$Z=-0.12h^{-1} \units{Mpc}$} (see
\cref{fig:cl6Maps_temp}). The line chosen is nearly perpendicular to
the shock front and CF and will thus allow us to probe the jumps in
values of the gas properties. At each point along the line we
average\footnotemark the gas properties in the plane perpendicular to
the line ($X-Z$ plane) on a scale of $40\units{kpc}$.
\footnotetext{Temperature and entropy averages are mass weighted.}

In \cref{fig:cl6Bore_prof} we show the profiles in the vicinity of the
shock and CF for the gas density, temperature, entropy pressure
and metallicity while the velocity component parallel ($v_y$) and
perpendicular ($v_x$ and $v_y$) to the reference line are shown in
\cref{fig:cl6Bore_vprof}. The locations of shocks and CFs are also
marked.
The profiles are in units of virial parameters (or combinations of
them): density is in units of $\vir{\rho} \equiv \delvir
\rho_{\mathrm{mean}}$, temperature in units of $\Tv$, entropy in units
of $\Tv/\vir{\rho}^{3/2}$ and pressure in units $\vir{\rho}
\Tv$. Metallicity is in units of $0.1 Z_\odot$ and the velocity
components are in units of $\Vv$.

It is common to separate the velocity at a discontinuity into two
components: one parallel and one perpendicular \emph{to the
  discontinuity}. In our treatment, the orientation of the
discontinuity is not defined, and instead we choose a reference line
which is roughly perpendicular to the discontinuity (see
\cref{fig:cl6Maps_temp}). We find it more precise to define the
velocity components with respect to the reference line, so the
velocity component parallel to the line is a good approximation to the
velocity component which is perpendicular to the discontinuities, and
vice-versa.
 
A closer examination of the profiles reveals that there are indeed two
shock fronts at the collision site, as evident by the jumps in
pressure, density, temperature and the velocity component
perpendicular to the shock front. Both propagating towards the right
in \cref{fig:cl6Bore}, which is upwards in \cref{fig:cl6Maps}. The
topmost shock is found at \mbox{$Y\sim 0.29 h^{-1} \units{Mpc}$} and
the second, more prominent shock at \mbox{$Y\sim 0.24 h^{-1}
  \units{Mpc}$}. The Mach number for the shocks can be deduced from
the jump in pressure (\equnp{rhP}) and the shock velocity can then be
calculated from \cref{eq:vshock1}, finding $u_{s,1}\simeq 1580
\units{km\,s^{-1}}$ ($\mach_1\simeq 1.5$) for the lower shock
(indicated by a red arrow in \cref{fig:cl6Maps}) and $u_{s,2}\simeq
840 \units{km\,s^{-1}}$ ($\mach_2\simeq 1.5$) for the upper shock
(indicted by a green arrow in \cref{fig:cl6Maps}). If a satellite was
indeed at the forefront of the stream penetration, the upper, weaker
shock may be a signature of the satellite reaching the apocentre,
while the second shock is the result of additional stream material
colliding with the ambient gas.

We thus have a situation in which the weaker shock might be overtaken
by the stronger one in $\sim 350\units{Myr}$, which might lead to the
formation of an additional CF when these two shocks merge
\citep{Birnboim2010}.

\begin{figure}
  \centering
  \subfloat[Temperature of the central region]{\label{fig:cl6Map06_temp}
    \includegraphics[height=7.25cm,keepaspectratio,bb=0 0 5.1in 4.42in]{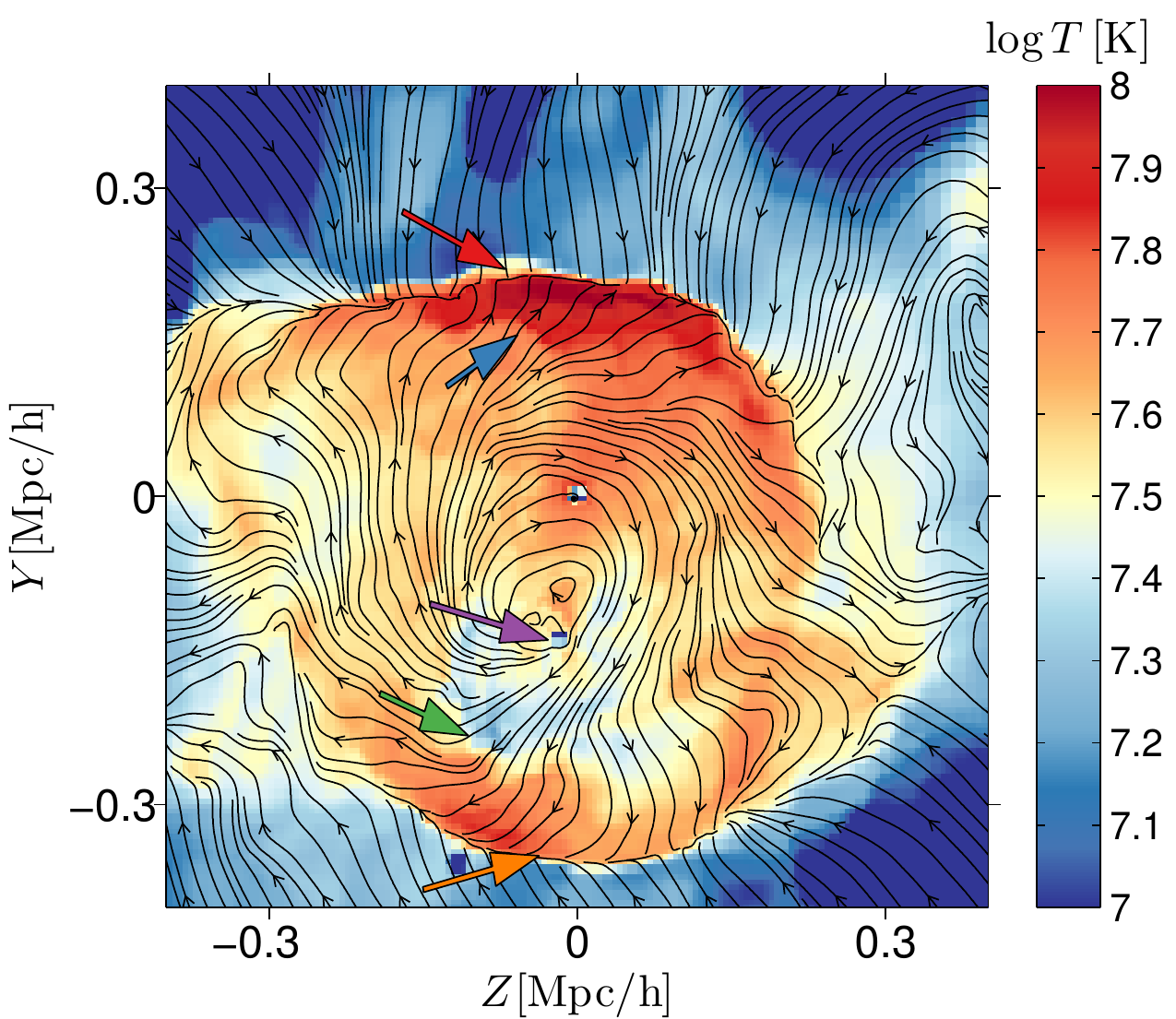}}\\
  \subfloat[Gas density in the central region]{\label{fig:cl6Map06_rho}
    \includegraphics[height=7.25cm,keepaspectratio,bb=0 0 5.5in 4.43in]{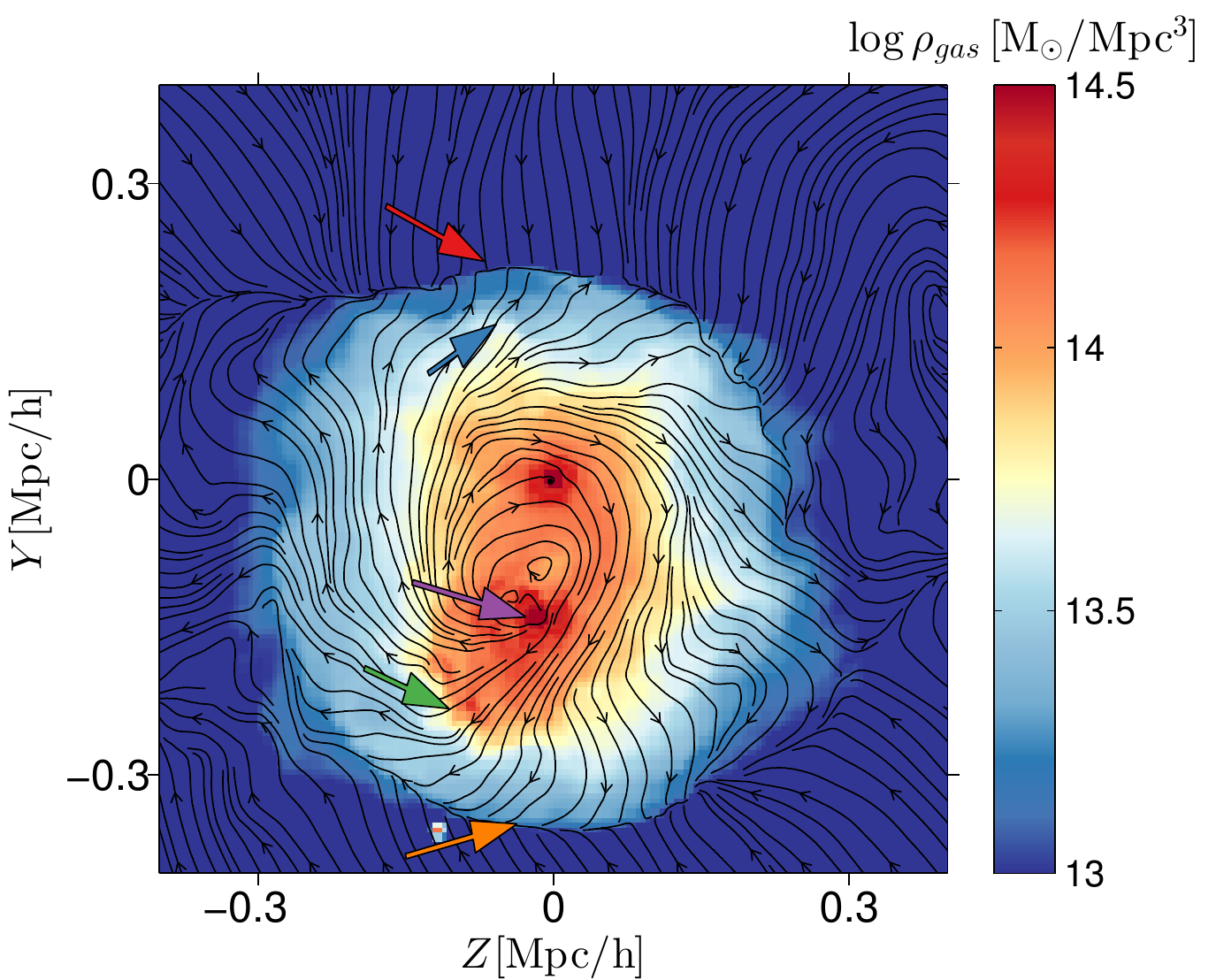}}
  \caption{The temperature \subrfig{cl6Map06_temp} and density
    \subrfig{cl6Map06_rho} of the gas in the central region of CL6 at
    \zeq{0.6}. We identify two shocks (red \& orange arrows)
    propagating into the streams as well as the CFs (blue \& green
    arrows). A large satellite is discernible just below the central
    galaxy marked by a purple arrow. The CF marked by the green arrow
    is very pronounced due to high metallicity gas stripped by the
    satellite. All maps are averaged over a slice of $125h^{-1}
    \units{kpc}$.}
  \label{fig:cl6Map06}
\end{figure}

Below the shock front, at a position of \mbox{$Y=0.19h^{-1} \units{Mpc}$}
we find the distinct CF with a sharp rise in temperature, with a
contrast of $q\simeq 2.4$ and an equal (to within a few per cent) drop in
density (along the positive $Y$ direction), which ensures pressure
equilibrium. The metallicity across the CF is roughly constant, making
it unlikely that the CF was formed due to gas stripping from a
satellite. The velocity component perpendicular to the CF is
continuous (unlike a shock) while a jump is found in the tangential
components (\cref{fig:cl6Bore_vprof}).

At a position of \mbox{$Y=0.13h^{-1} \units{Mpc}$} we find another
feature we suspect of being a shock, though the gradient along the
line is not as sharp as the other shock features, most likely due to
inclination of the shock front with respect to the reference line
(\cref{fig:cl6Maps_press}). If this is indeed a shock, it is
propagating downwards and we find that the idealized scenario
presented in \rfsec{coldFronts} has been realized in the simulation,
with two shocks moving in opposite directions\footnotemark, with a CF
in between the two. The Mach number and velocity of this shock are
$\mach_3\simeq1.3$ and $u_{s,3}\simeq -280\units{km\,s^{-1}}$,
respectively.  \footnotetext{The velocities of the shocks are both
  positive to an outside observer, but the shocks are receding from
  each other.}

In \cref{fig:cl6Map06_flux} we show the mass inflow rate of the
cluster CL6 at an earlier epoch of \zeq{0.6}, with the temperature and
density in the centre of the cluster at that time shown in
\cref{fig:cl6Map06}. The inflowing streams existed back then in a
similar configuration, with some notable differences. At \zeq{0.6} we
find two streams, one flowing from the top (which at \zeq{0} was
stopped at the virial radius) and another from the bottom towards the
centre where two large shocks can be seen propagating back into the
streams, marked by red and orange arrows in \cref{fig:cl6Map06_temp},
in a configuration similar to the ideal 1D model discussed in
\rfsec{coldFronts}. CFs can also be identified in the pictures (as indicated 
by blue and green arrows).

A large satellite can be identified just below the central galaxy, and
the cooler gas found just above the lower CF (marked by the green
arrow) contains high-metallicity gas stripped from the satellite. The
formation of the lower CF is therefore due at least in part to the gas
removed from the satellite galaxy.


This is hardly surprising since in the period of $5.7\units{Gyr}$
between the two epochs, much can happen in a dynamically unrelaxed
cluster such as this one. In a detailed analysis of the streams in the
simulated cluster suite \citet{Zinger2016} find that the penetration of
the streams may change with streams either penetrating deeper or being
pushed out over time. With the changes of stream penetration, new
instances of collisions between streams and the surrounding gas can
lead to the formation of shocks and CFs.

\subsection{Colliding Streams in CL107}\label{sec:cl107}
In this section we present an additional example of a stream collision leading to the
formation of shocks and CFs at the collision site.

\begin{figure}
  \centering \subfloat[Gas Density]{\label{fig:cl107Map8_gas}
    \includegraphics[width=9 cm,keepaspectratio,bb=0 0 5.33in 4.43in]{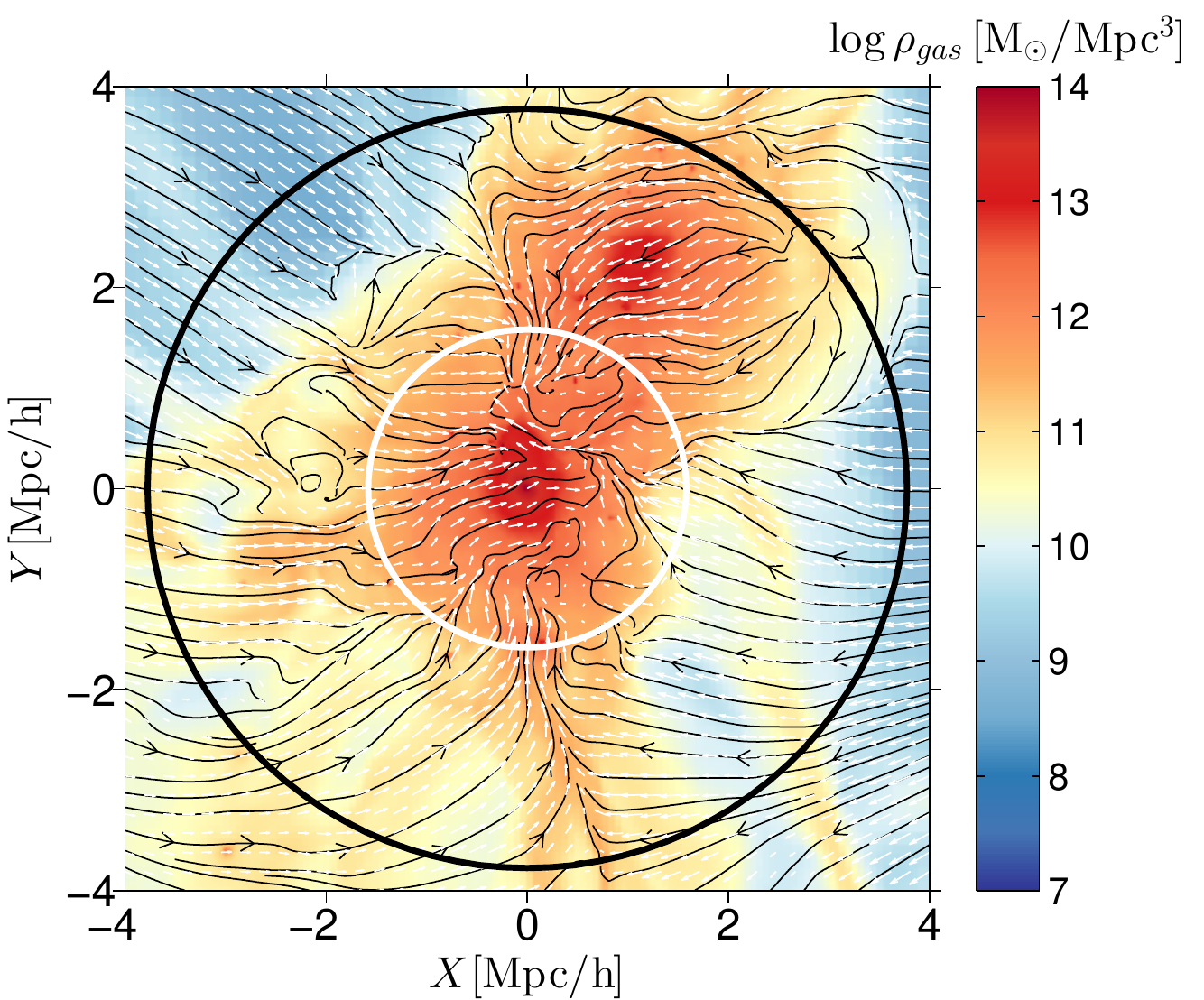}}\\
  \subfloat[Dark matter Density]{\label{fig:cl107Map8_dm}
    \includegraphics[width=9 cm,keepaspectratio,bb=0 0 5.38in 4.43in]{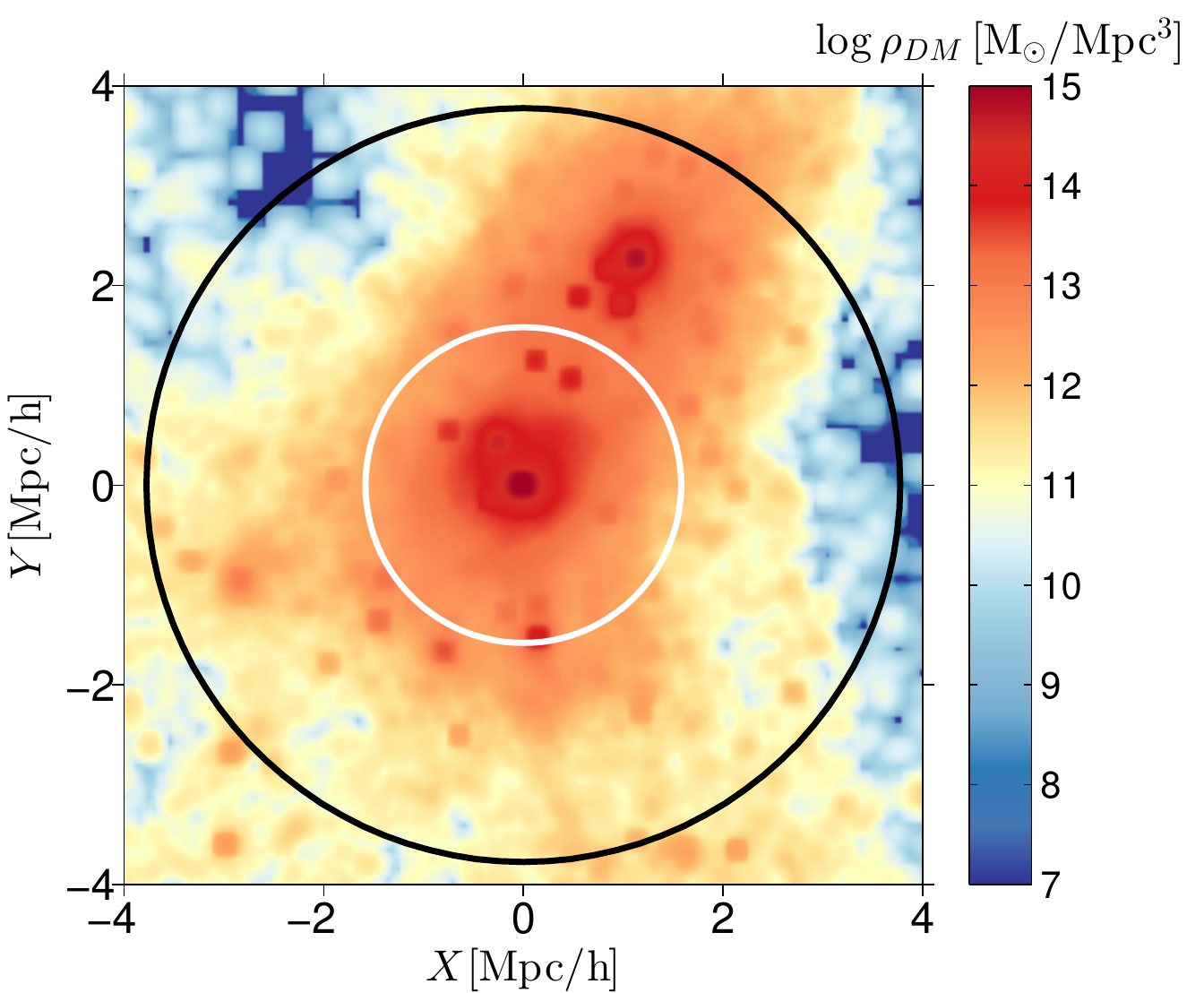}}
  \caption{Gas \subrfig{cl107Map8_gas} and dark matter density
    \subrfig{cl107Map8_dm} of the cluster CL107 at \zeq{0}.  The maps
    and lines are created in the same way as those \cref{fig:cl6Map8}.
    An additional system, found in the top right quadrant, will soon
    cross the virial radius of the cluster. }
  \label{fig:cl107Map8}
\end{figure}

\begin{figure*}
  \centering
  \subfloat[Velocity]{\label{fig:cl107Maps_vel}
    \includegraphics[height=6.3cm,keepaspectratio,bb=0 0 5.35in 4.42in,trim=0.70in 0.425in -0.06in  0.06in, clip]{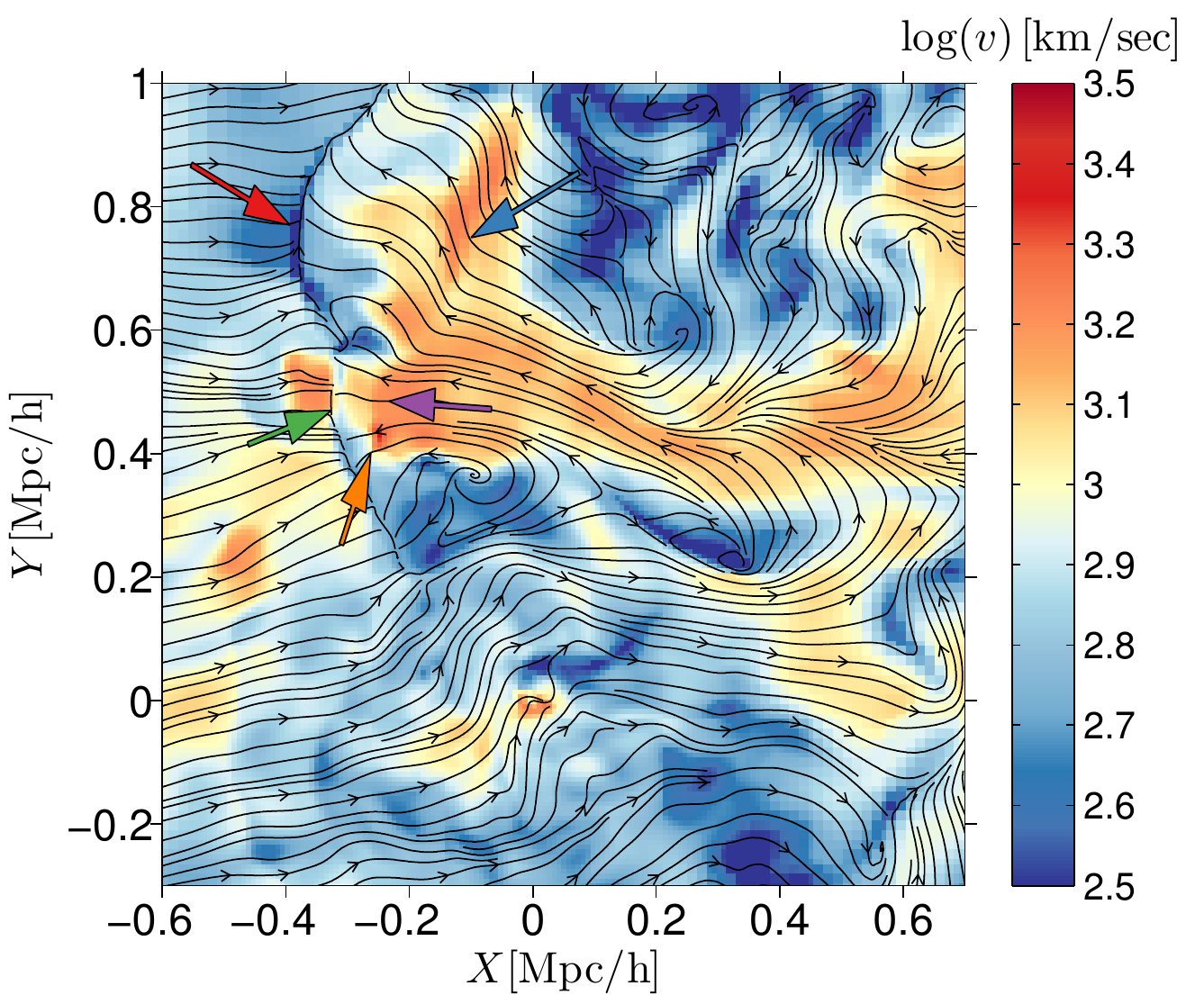}} 
  \subfloat[Metallicity]{\label{fig:cl107Maps_zmet}
    \includegraphics[height=6.3cm,keepaspectratio,bb=0 0 5.19in 4.42in,trim=0.70in 0.425in -0.22in  0.06in, clip]{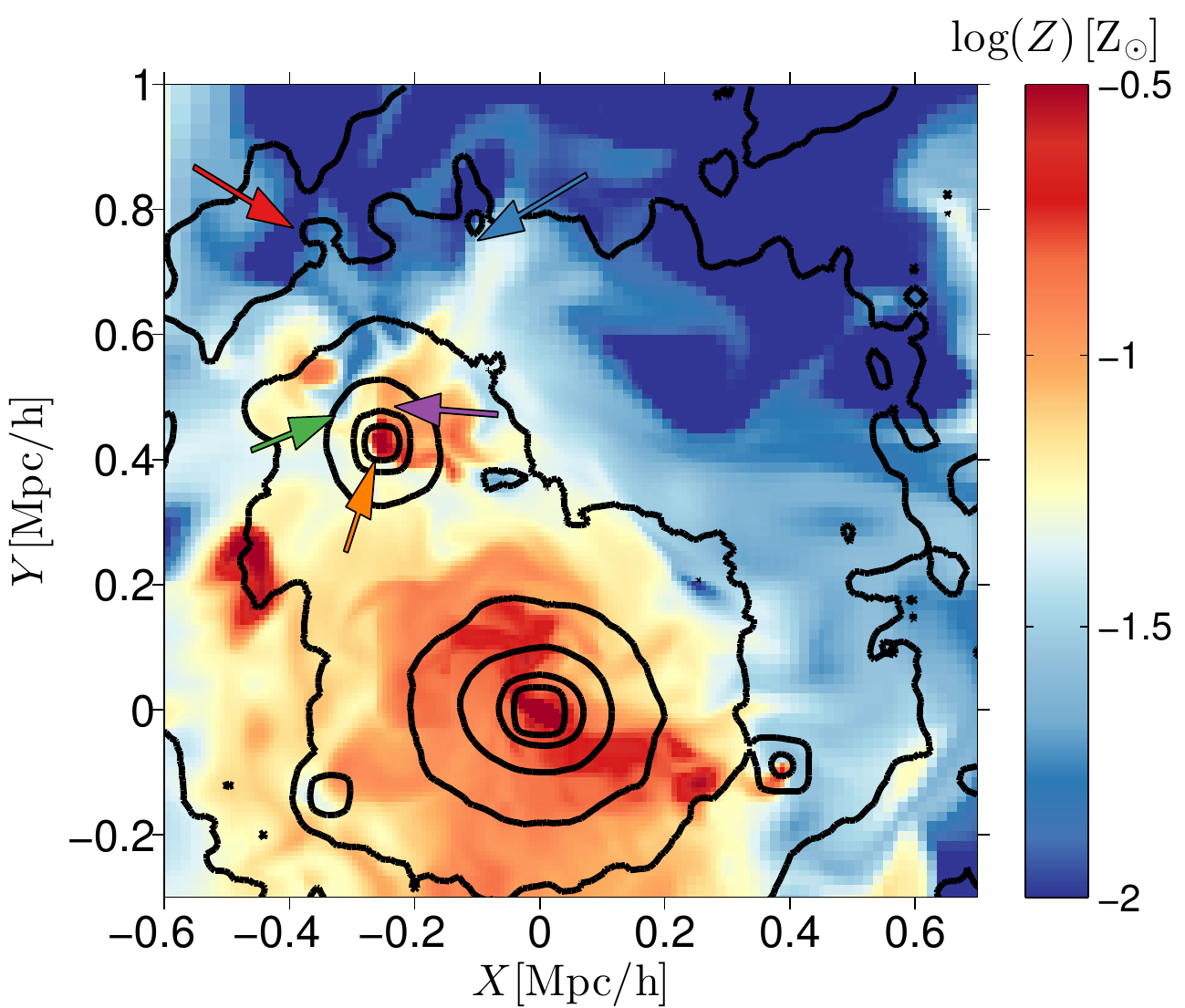}}\\
  \subfloat[Entropy]{\label{fig:cl107Maps_ent}
    \includegraphics[height=6.3cm,keepaspectratio,bb=0 0 5.36in 4.43in,trim=0.70in 0.42in -0.05in  0.06in, clip]{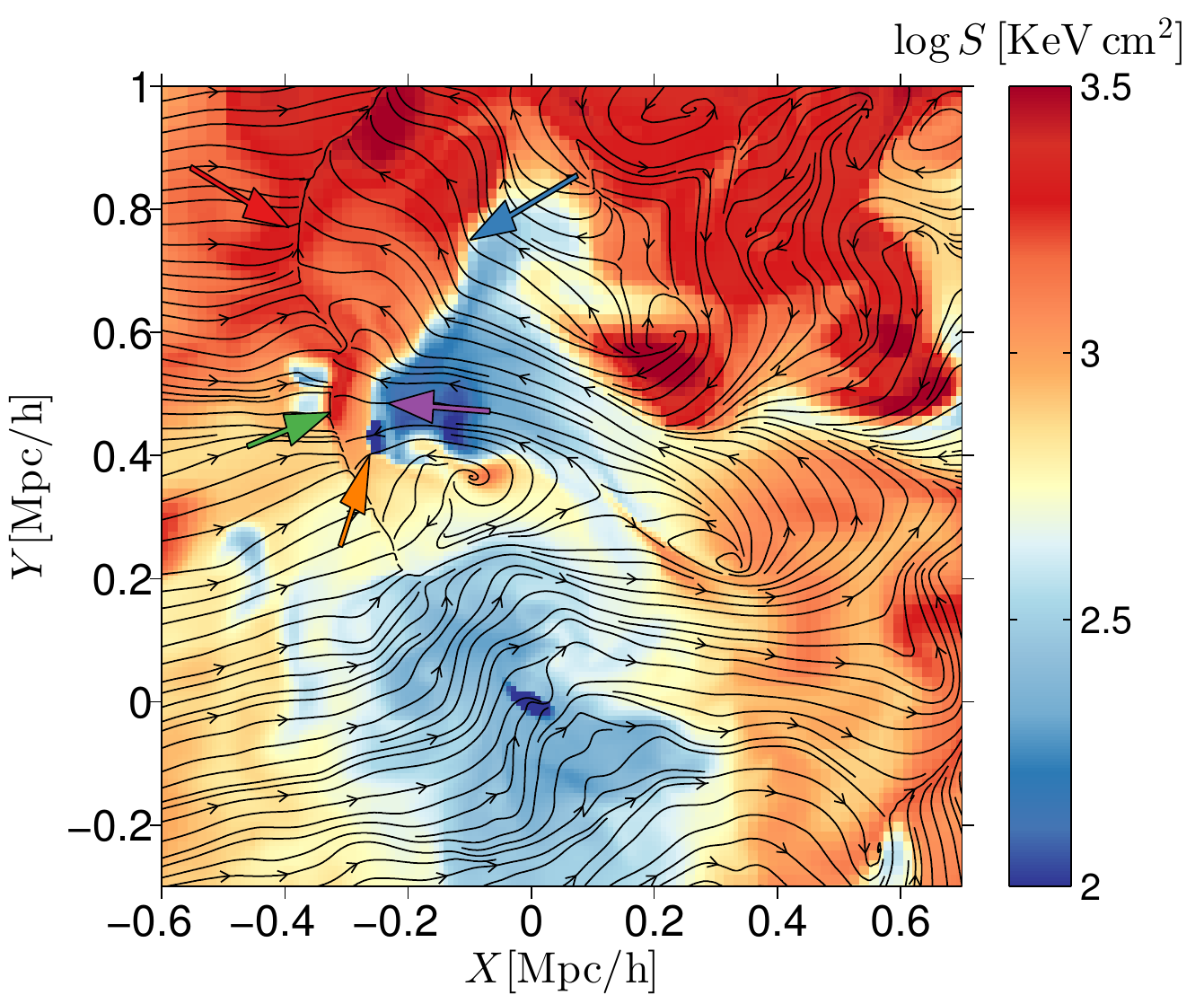}} 
  \subfloat[Temperature]{\label{fig:cl107Maps_temp}
    \includegraphics[height=6.3cm,keepaspectratio,bb=0 0 5.1in 4.42in,trim=0.70in 0.42in -0.31in  0.06in, clip]{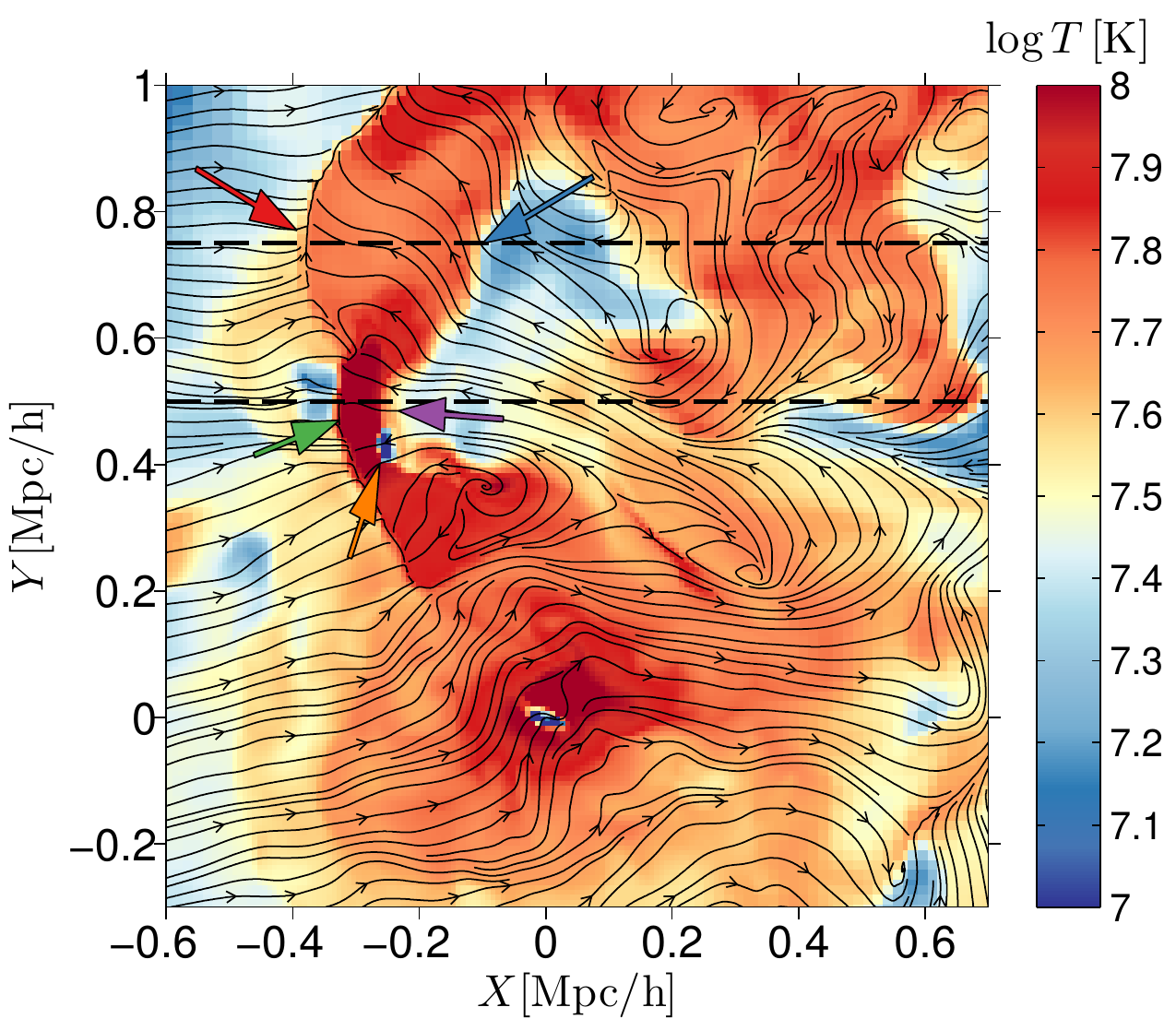}}\\
  \subfloat[Pressure]{\label{fig:cl107Maps_press}
    \includegraphics[height=6.3cm,keepaspectratio,bb=0 0 5.19in 4.42in,trim=0.70in 0.42in -0.22in  0.06in, clip]{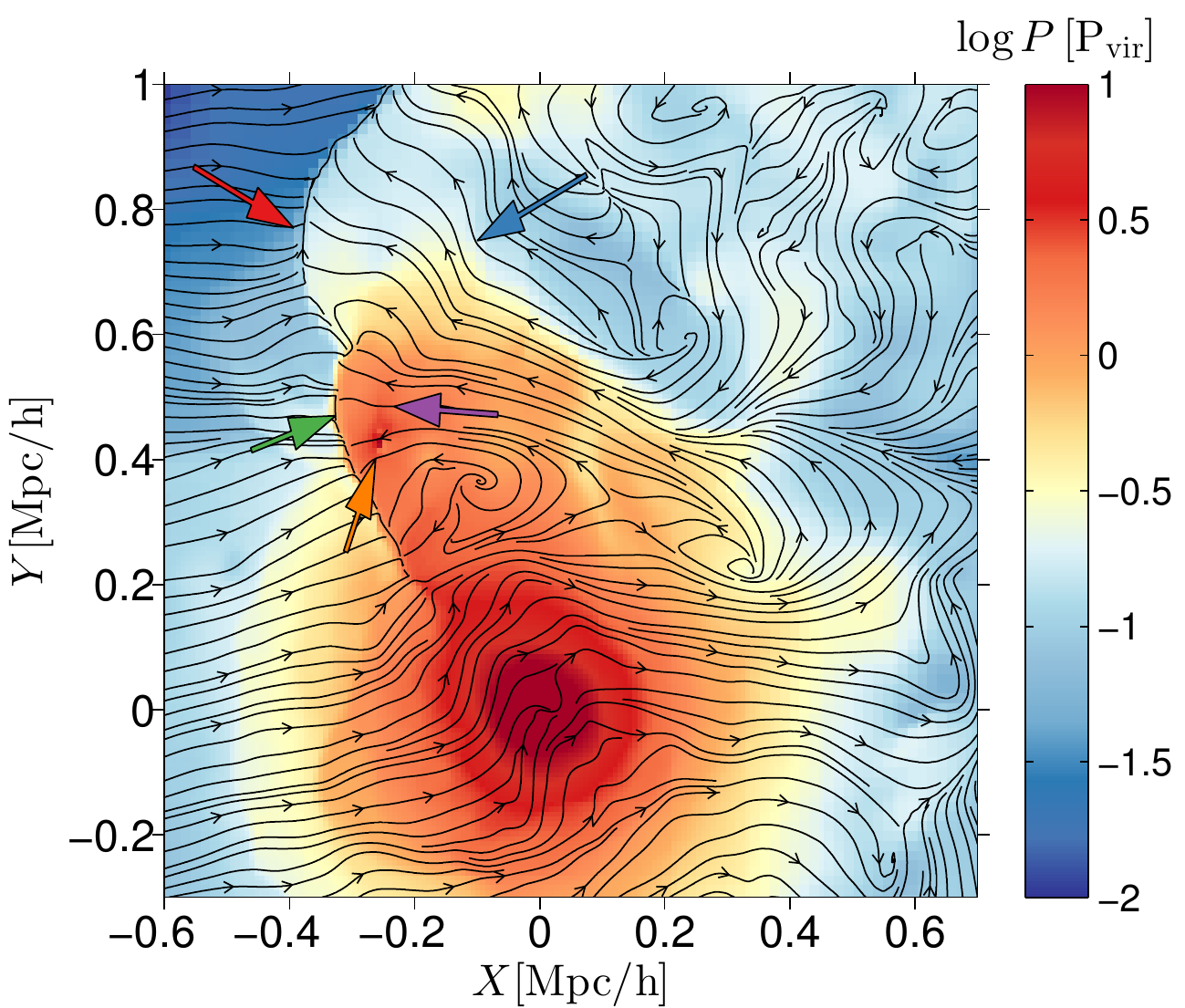}} 
  \subfloat[Gas Density]{\label{fig:cl107Maps_rho}
    \includegraphics[height=6.3cm,keepaspectratio,bb=0 0 5.5in 4.43in,trim=0.70in 0.42in -0.09in  0.06in, clip]{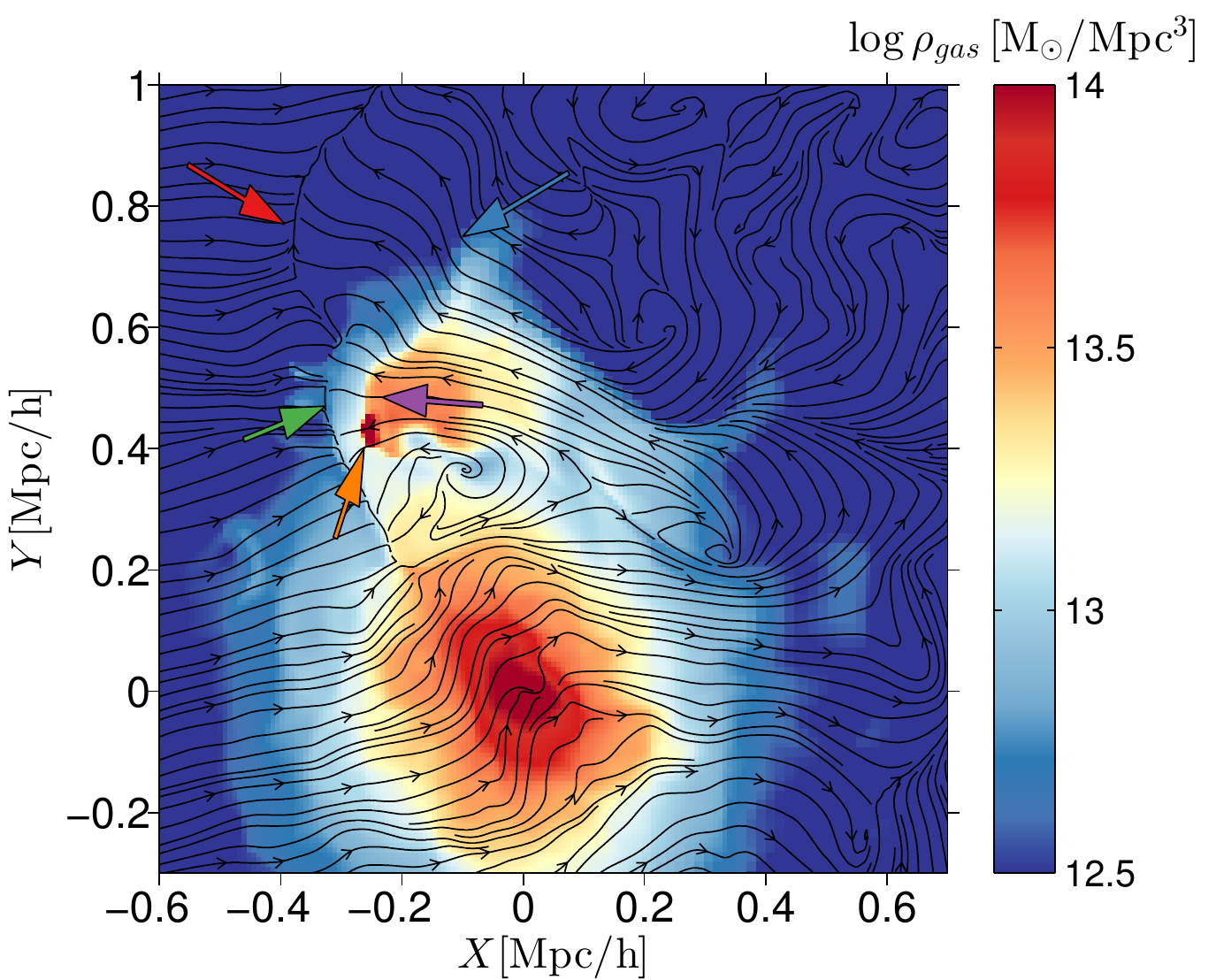}}
  \caption{The collision zone between the two streams in CL107 at
    \zeq{0} is shown in detail in a box of size $1.3h^{-1}
    \units{Mpc}$ centered around $(0.05,0.35)$ on the
    $X-Y$ plane.  The velocity \subrfig{cl107Maps_vel} metallicity
    \subrfig{cl107Maps_zmet} entropy \subrfig{cl107Maps_ent}
    temperature \subrfig{cl107Maps_temp} pressure
    \subrfig{cl107Maps_press} and gas density \subrfig{cl107Maps_rho}
    are all averaged over a slice of $25h^{-1} \units{kpc}$. Dark
    matter density contours (black) are shown in
    \subrfig{cl107Maps_zmet} (averaged over $50h^{-1} \units{kpc}$).
    Streamlines represent the velocity field.  Areas of interest are
    marked by colored arrows as follows: the red \& green arrows mark
    two shock fronts. The CFs are marked by blue \& purple arrows. An
    orange arrow points to the satellite galaxy. The dashed horizontal
    lines shown in \subrfig{cl107Maps_temp} mark the values used to
    follow the profiles of gas properties shown in
    \cref{fig:cl107BoreTop,fig:cl107BoreBottom}.}
  \label{fig:cl107Maps}
\end{figure*}

\begin{figure}
  \centering
  \subfloat[Thermodynamic Properties]{\label{fig:cl107BoreBottom_prof}
    \includegraphics[width=8.5cm,keepaspectratio,bb=0 0 5.28in 4.18in,trim=0.0in 0.0in 0.0in  0.0in, clip]{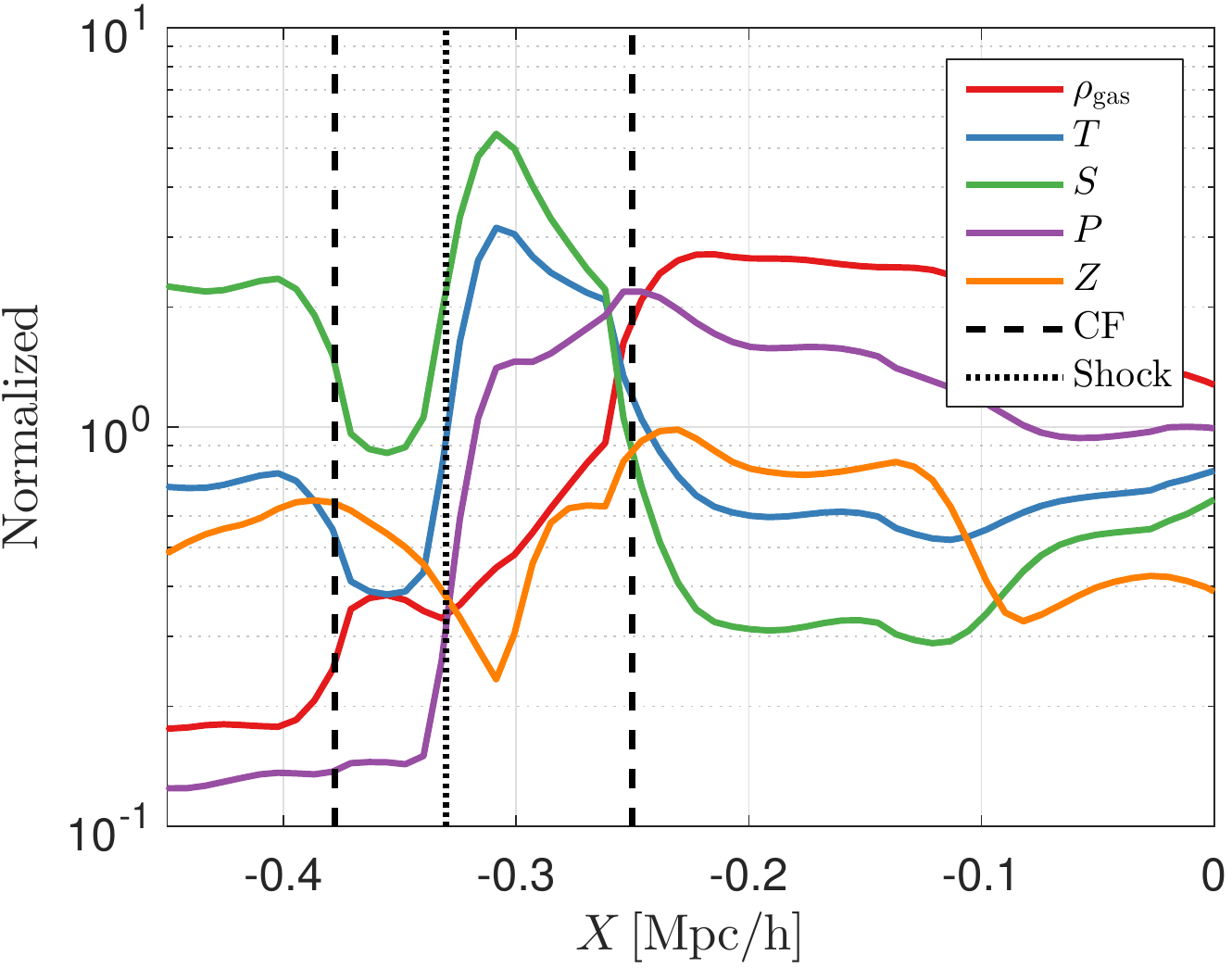}}\\
  \subfloat[Velocity Components]{\label{fig:cl107BoreBottom_vprof}
    \includegraphics[width=8.5cm,keepaspectratio,bb=0 0 5.24in 4.15in,trim=0.0in -0.015in -0.04in  -0.015in, clip]{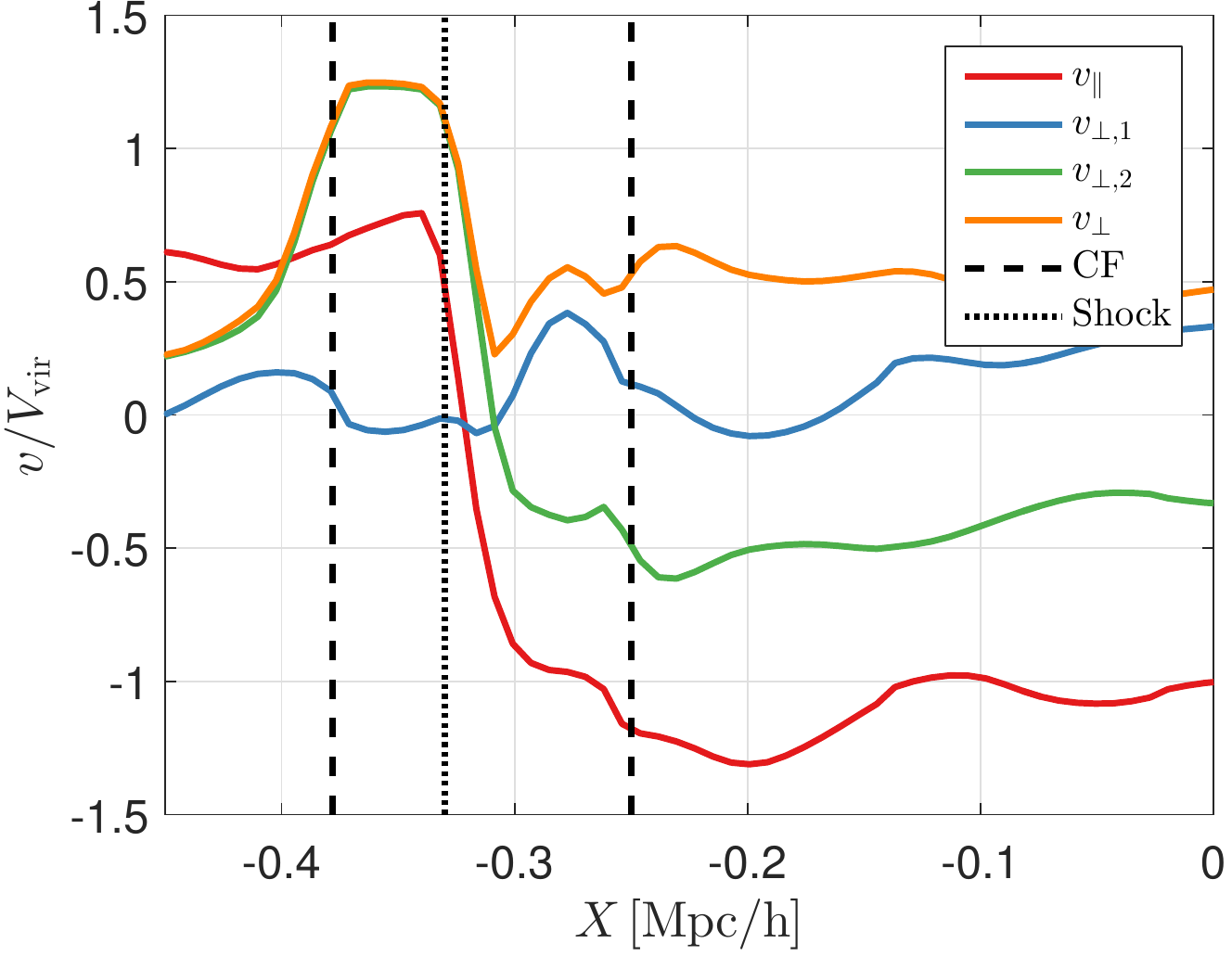}}
  \caption{Profiles of the gas properties in CL107 at \zeq{0} along a
    line in the $X$ direction, perpendicular to the shock generated
    near the satellite and CF (green and purple arrows in
    \cref{fig:cl107Maps}) at \mbox{$Y=0.5h^{-1} \units{Mpc}$} (see
    \cref{fig:cl107Maps}). Gas properties and line types are same as
    those in \cref{fig:cl6Bore}.  Locations of shock fronts and CFs
    are also marked by black dotted and black dashed lines,
    respectively. The CF to the right of the shock is formed by
    stripping of gas the infalling satellite. The CF to the left of
    the shock is most likely a relic, which formed by the stream
    collision prior to the arrival of the satellite. }
  \label{fig:cl107BoreBottom}
\end{figure}
\begin{figure}
  \centering
  \subfloat[Thermodynamic Properties]{\label{fig:cl107BoreTop_prof}
    \includegraphics[width=8.74cm,keepaspectratio,bb=0 0 5.36in 4.18in,trim=0.0in 0.0in -0.08in  0.0in, clip]{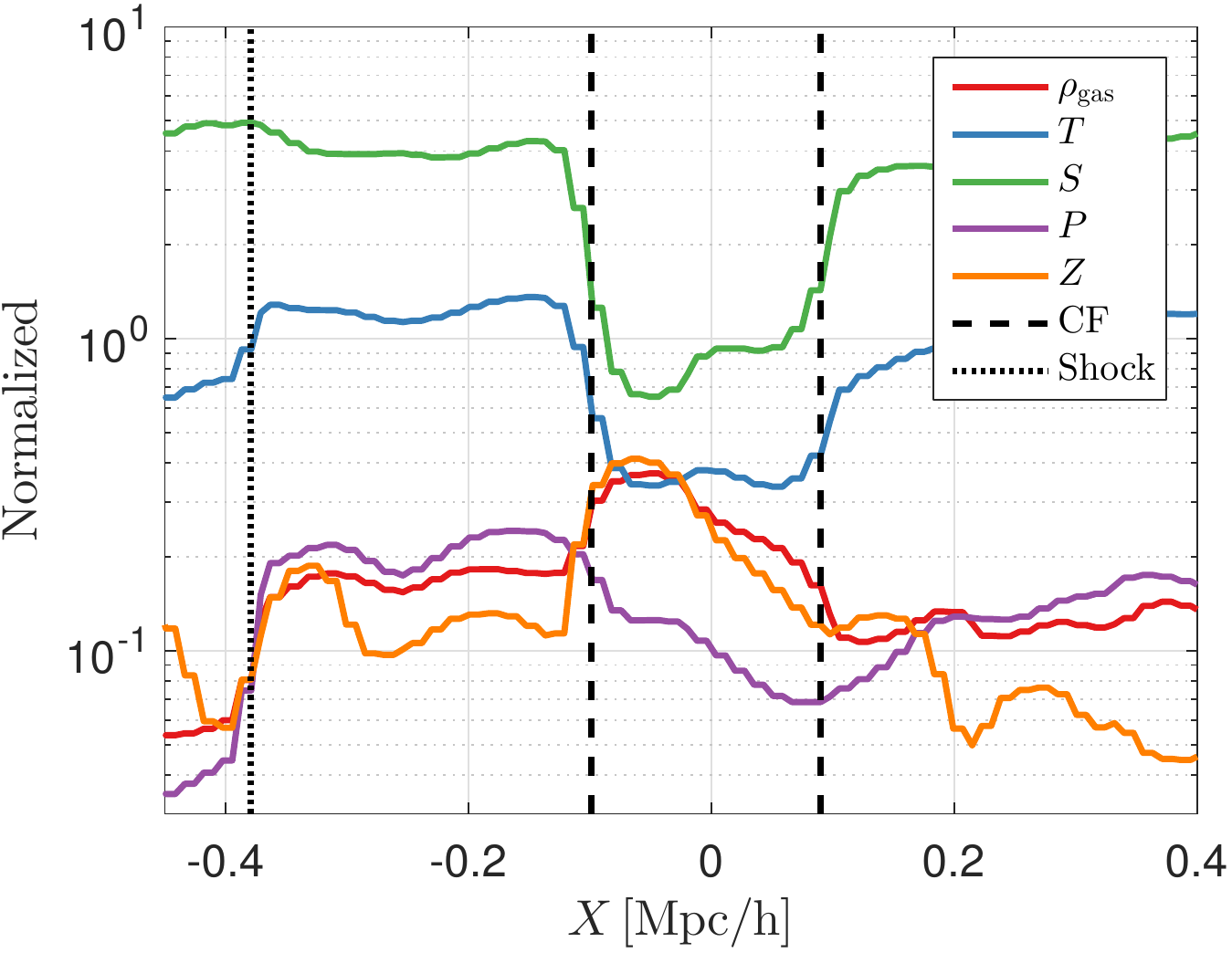}}\\
  \subfloat[Velocity Components]{\label{fig:cl107BoreTop_vprof}
    \includegraphics[width=8.74cm,keepaspectratio,bb=0 0 5.32in 4.15in,trim=0.0in -0.015in -0.12in  -0.015in, clip]{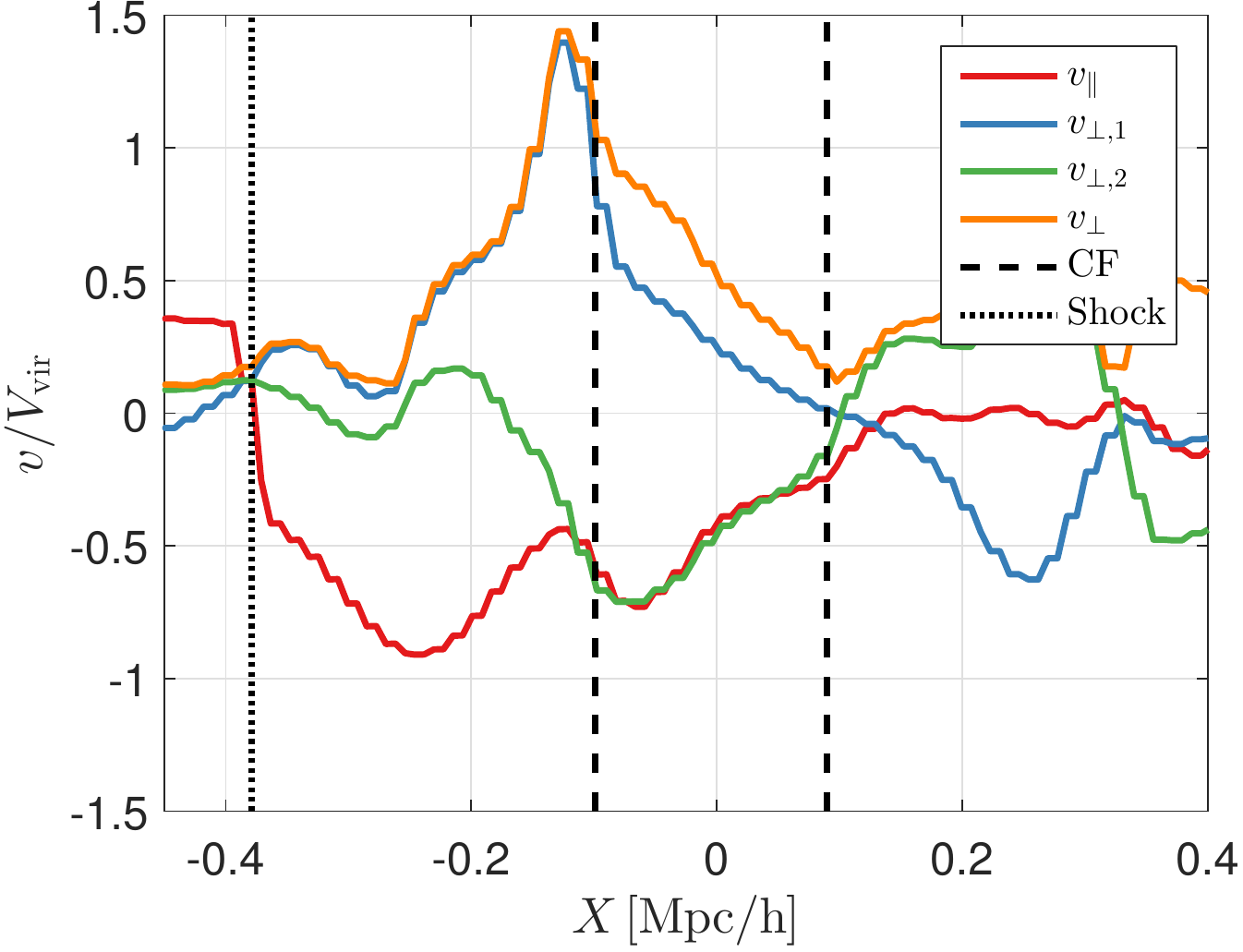}}
  \caption{Profiles of the gas properties \subrfig{cl107BoreTop_prof}
    and velocity components \subrfig{cl107BoreTop_vprof} in CL107
    along a line in the $X$ direction, perpendicular to the shock and
    CF site formed by the stream collision (red and blue arrows in
    \cref{fig:cl107Maps}) at \mbox{$Y=0.75h^{-1} \units{Mpc}$} (see
    \cref{fig:cl107Maps}). Gas properties and line types are the same
    as in \cref{fig:cl107BoreBottom}. The sharp change in metallicity
    across the CF found on the left shows the enriched gas stripped
    from the satellites that are partially responsible for the
    formation of the CF.}
  \label{fig:cl107BoreTop}
\end{figure}

The cluster CL107 has a virial mass of $\Mv=6.6\times 10^{14} \msun$
and a virial radius of $\Rv=2.26\units{Mpc}$ at \zeq{0} (see
\cref{tab:clusterProperties}).  \cref{fig:cl107Map8} shows the gas and
dark matter density of the cluster on a scale of several $\Rv$. The
cluster is situated on a junction of 3 large scale filaments: one from
the top right corner, one from the left and one from the bottom. Along
the top right filament a merging cluster is poised to cross the virial
radius in the near future. The accretion shock extends to $\sim 2.8
\Rv$ \citep{Zinger2016b}.

A closer examination of these maps reveals that the gas from the
prominent stream enters with the gas velocity of $\sim
1000\units{km\,s^{-1}}$ from the top right, upon entering within the
virial radius of the cluster, and bends its motion to the negative $X$
direction, leading to the collision with another stream coming from
the left in the inner region of the cluster.

In \cref{fig:cl107Maps} we zoom in on the area of stream collision,
and once again highlight important features with colored arrows. The
shock front is roughly $750h^{-1} \units{kpc}$ long, clearly visible
in the temperature, pressure and density maps
(\cref{fig:cl107Maps_temp,fig:cl107Maps_press,fig:cl107Maps_rho}). Closer
examination, especially in the pressure map
(\cref{fig:cl107Maps_press}), reveals that the shock front is
comprised of two distinct shocks. The top one (red arrow) originates
with the collision of the two gas streams, and the bottom one (green
arrow) is due to a satellite galaxy.

The satellite galaxy, marked by the orange arrow, can be seen as a
cold dense spot just behind the shock front, and is especially evident
in the metallicity map (\cref{fig:cl107Maps_zmet}), which also shows
the dark matter density distribution. One can clearly see the highly
enriched gas being stripped from the galaxy. The gas behind the top
shock front (red arrow) can be seen to be nearly devoid of metals, and
coupled with the absence of a dark matter sub-halo in that region,
leading to the conclusion that this shock front did not originate from
a satellite. Finding a satellite along the inflowing gas stream is
expected since the streams mark the preferred direction of accretion
into the cluster.

A CF is naturally found behind the two shock fronts and it too extends
over quite a large distance (blue and purple arrows). When examining
the density and metallicity of the `cold' side of the CF, it seems
that most of it may have originated in the merging satellite. It is
thus unclear whether the CF formed naturally behind the collision
shock, or that the CF is the interface between the hot post shock gas
and the cooler, denser gas stripped from the satellite. It is of
course possible that the two processes are involved.

As before we examine the profiles of various gas properties along two
horizontal lines indicated in \cref{fig:cl107Maps_temp}, which are
nearly perpendicular to the shock and CF.  The top line located at
\mbox{$X=0.75 h^{-1} \units{Mpc}$} goes through the upper, stream
collision shock (\cref{fig:cl107BoreTop}), while the other line
located at \mbox{$X=0.5 h^{-1} \units{Mpc}$} passes through the lower,
merger induced shock (\cref{fig:cl107BoreBottom}).

Along the bottom line of \cref{fig:cl107Maps_temp}, we find in
\cref{fig:cl107BoreBottom} the merger induced shock at \mbox{$X=-0.33
  h^{-1} \units{Mpc}$} moving with the gas velocity of
$u_{s}\simeq-950\units{km\,s^{-1}}$ ($\mach\simeq2.8$).  The sign
reversal of the parallel velocity component along the shock is a
testament to the dramatic nature of head-on collision between the two
streams.  Behind the shock at \mbox{$X=-0.25 h^{-1} \units{Mpc}$} a CF
can be identified with density/temperature contrast of $q\simeq 3.1$.
Another, smaller CF can be seen, just ahead of the shock wave at
\mbox{$X=-0.38 h^{-1} \units{Mpc}$}, with a density/temperature
contrast of $q\simeq 2$. In both cases, the temperature and density
contrasts are equal to within \perc{5}. While the first CF can be a
result of the stream collision, the origin of the second one is not so
clear, and may be a relic from an earlier event.

For the top line the shock is located at \mbox{$X=-0.38h^{-1}
  \units{Mpc}$} and moving with the gas velocity
$u_{s}\simeq-1300\units{km\,s^{-1}}$ ($\mach\simeq1.9$).  Two CFs can
be seen at \mbox{$X=-0.1 h^{-1} \units{Mpc}$} and \mbox{$X=0.09 h^{-1}
  \units{Mpc}$}. The contrasts for the density and temperature are not
equal as in the other examples with a ratio of temperature to density
contrasts, which we designate as $\eta$
\begin{equation}\label{eq:qRatio}
\eta\equiv q_T q_\rho= \frac{T_1}{T_2} \frac{\rho_1}{\rho_2} ,
\end{equation}
and takes the values of $\eta_1=1.8$ and $\eta_2=0.8$ for the
two CFs.

This shows that there is something different about the way these CF
developed compared to the CF found in CL6. Unlike the other examples
(\cref{fig:cl6Bore,fig:cl107BoreBottom}), in which the metallicity
profile showed little variation, the metallicity of the gas between
the two CFs is markedly higher. In addition, the feature on the right
while resembling a CF with a drop in temperature and a rise in density
also has a drop in pressure, signifying that it may not be a CF but
rather a region of colder and denser gas pushing against a warmer and
more dilute region.

The velocity along the line (red line in \cref{fig:cl107BoreTop_vprof}
shows that the cooler gas is indeed travelling towards the left
slightly faster than the gas beyond the interface. There is also a
sharp change in the perpendicular velocity component across the
CFs. This is in accordance with our earlier conclusion that the CF are
due, at least in part, to the presence of gas which has been stripped
from an infalling galaxy.

\begin{figure}
  \centering
  \includegraphics[width=8.5cm,keepaspectratio,bb=0 0 5.25in 4.07in]{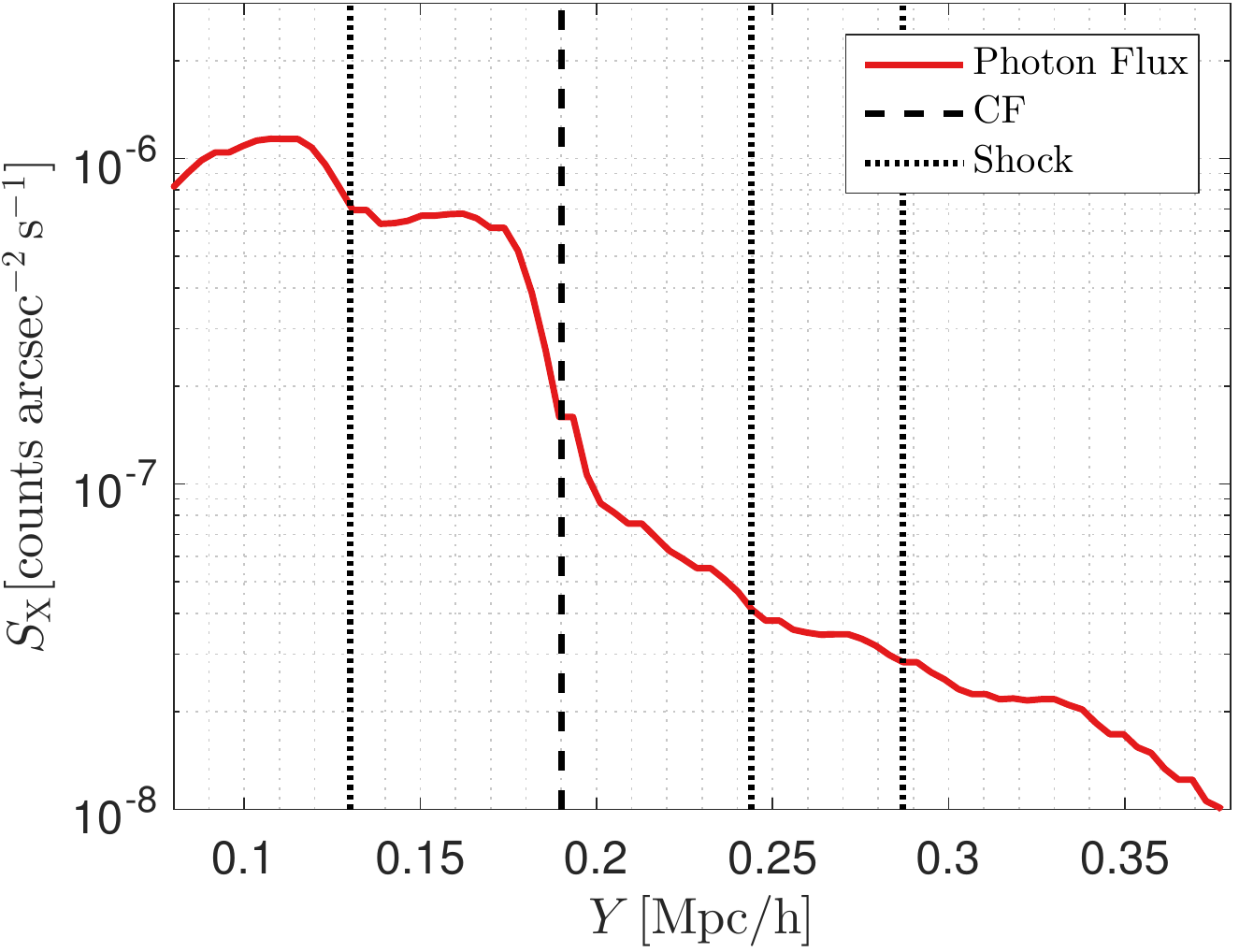}
  \caption{The X-ray surface brightness in the
    $0.5\textrm{--}2\units{keV}$ range as estimated from the
    simulation data of CL6 along a line in the $Y$ direction (and
    $X=0$), perpendicular to the shock and CF, at \mbox{$Z=-0.12h^{-1}
      \units{Mpc}$} (see \cref{fig:cl6Maps_temp}). Locations of shock
    fronts and CFs are also marked by black dotted and black dashed
    lines, respectively. The X-ray flux is summed over
    $1\units{Mpc\,h^{-1}}$ along the $X$ direction. The CF feature is
    clearly discernible and is potentially detectable with current
    observational instruments. }
  \label{fig:cl6XrayBore}
\end{figure}

\begin{figure}
  \centering \subfloat[Bottom CF in
    CL107]{\label{fig:cl107XrayBoreBottom}
    \includegraphics[width=8.74cm,keepaspectratio,bb=0 0 5.29in 4.16in,trim=0.0in 0.0in -0.08in 0.0in, clip]{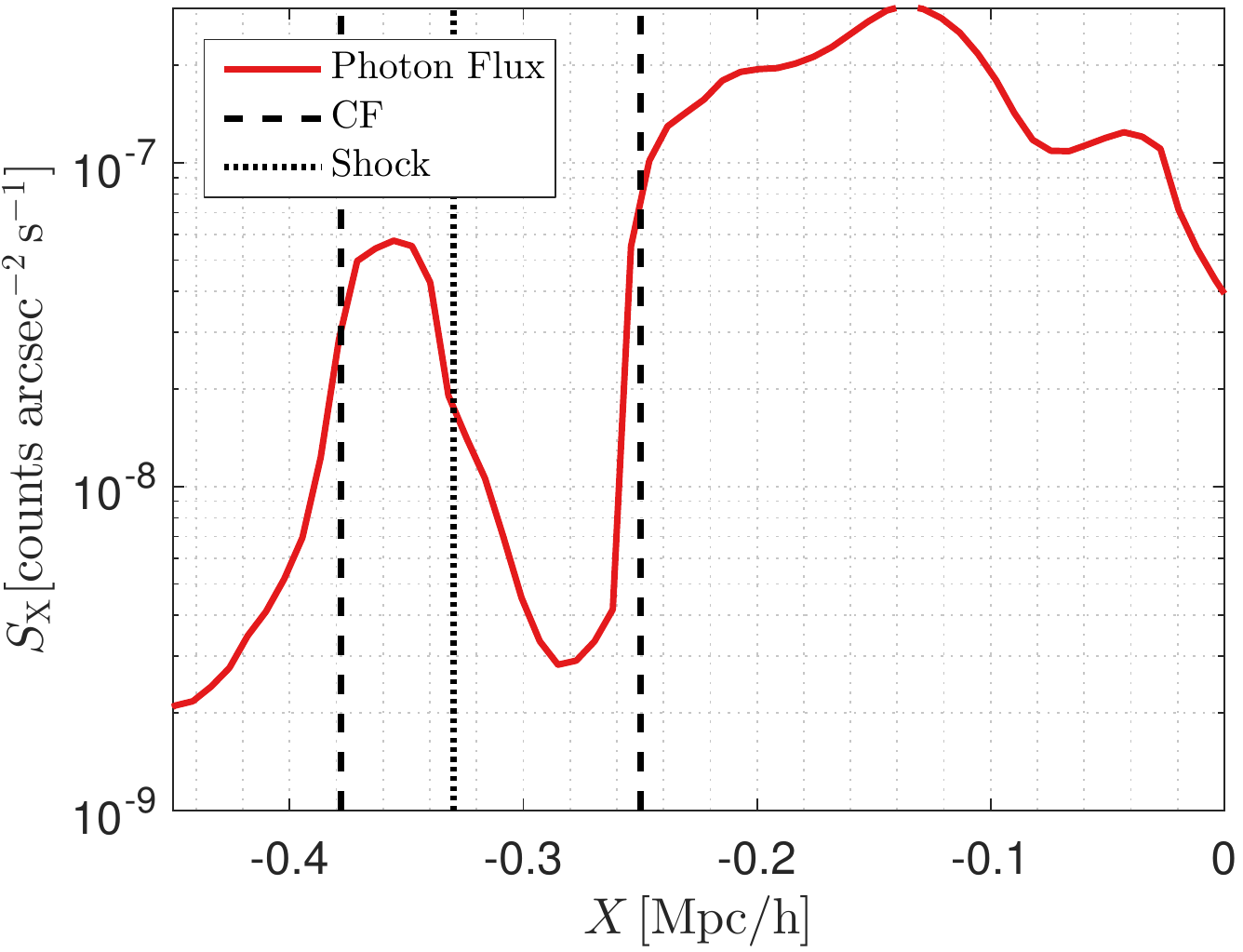}}\\
  \subfloat[Bottom CF in CL107]{\label{fig:cl107XrayBoreTop}
    \includegraphics[width=8.74cm,keepaspectratio,bb=0 0 5.36in 4.16in,trim=0.0in 0.0in -0.08in 0.0in, clip]{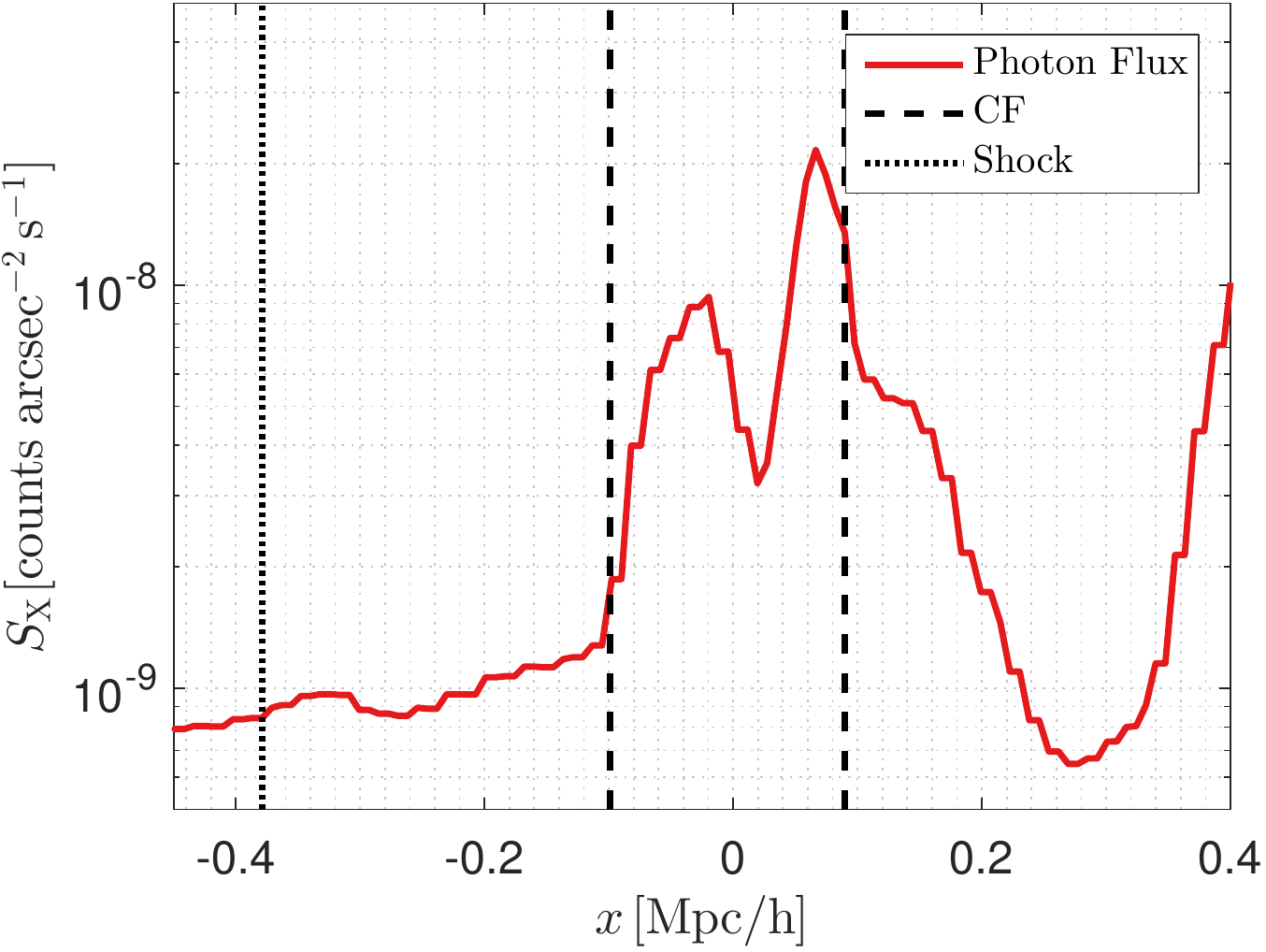}}
    \caption{The X-ray photon flux in the $0.5\textrm{--}2\units{keV}$ range as
      estimated from the simulation data of CL107 along a two lines in
      the $X$ direction at \mbox{$Y=0.5h^{-1} \units{Mpc}$}
      \subrfig{cl107XrayBoreBottom} and \mbox{$Y=0.75h^{-1}
        \units{Mpc}$} \subrfig{cl107XrayBoreTop}, perpendicular to the
      shock and CF site formed by the stream collision (see
      \cref{fig:cl107Maps_temp}). The X-ray flux is summed over
      $2\units{Mpc\,h^{-1}}$ along the $Z$ direction. The locations of
      shock fronts and CFs are marked by black dotted and black dashed
      lines, respectively. The low surface brightness in this regions
      most likely places these features beyond current observational
      capabilities.}
  \label{fig:cl107XrayBore}
\end{figure}

\section{Observational Detectability of Cold Fronts}\label{sec:detect}
To predict the detectability of the CFs and shocks discussed above, we
compute the X-ray surface brightness in the rest-frame energy band $E
= [0.5, 2.0]\units{keV}$ in the regions of interest of the clusters as
would be observed by current instruments of the {\em Chandra}
observatory.

Specifically, for each gas cell in the simulation, we compute the
X-ray emissivity per unit volume within $\epsilon(E)$ using the
\textsc{APEC} plasma code \citep[ver. 2.0.2][]{Smith2001, Foster2012}:
\begin{equation}
\epsilon(E) = n^2 \Lambda (E, T, Z, z),
\end{equation}
where $\Lambda (E, T, Z, z)$ is the specific cooling function from the
\textsc{APEC} code, $n$, $T$, $Z$ are the gas number density,
temperature, and metallicity for each cell respectively, and $z$ is
the observed redshift of the cluster which we set to \zeq{0.06}. For
the energy band we are considering here, the emissivity is only
weakly dependent on temperature and metallicity and is effectively a
measure of the number density \mbox{$(\epsilon\propto n^2)$}.

The observed X-ray photon count per unit time $f_x$ for each gas cell in the 
$E = [0.5, 2.0]\units{keV}$ energy band is computed as
\begin{equation}
f_x = \int^{2.0}_{0.5} \frac{\epsilon (E) l_c^3 }{4\mathrm{\pi} D_{\rm L}(z)^2} {\rm ARF}(E)dE,
\end{equation}
where $l_c$ is the cell-size, $D_{\rm L}(z)$ is the luminosity distance
and ${\rm ARF}(E)$ is the effective area of the {\em Chandra} ACIS-I
detector. We then sum up the contributions along the line of sight
and find the surface brightness $S_x$ by dividing by the angular area
subtended by the simulation cell on the sky
\begin{equation}
  S_x=\frac{f_x}{\left(l_c/D_{\rm A}(z)\right)^2},
\end{equation}
where $D_{\rm A}(z)$ is the angular diameter distance.

In \cref{fig:cl6XrayBore} we show the X-ray surface brightness profile
for the CF and shocks studied in CL6 along the same line used for the
profiles in \cref{fig:cl6Bore} (see also \cref{fig:cl6Maps_temp}). In
comparing the surface brightness profiles to the density profiles in
\cref{fig:cl6Bore_prof} it is important to bear in mind that former
are summed along the line of sight of the simulation data whereas the
latter are {\em local} quantities.

The shock features are not very prominent but the drop in surface
brightness associated with the CF is seen quite clearly. In terms of
detectability, the surface brightness of the CF is of order
$10^{-7}\units{counts\,arcsec^{-2}\,sec^{-1}}$ which is within the
observational capabilities of deep {\em Chandra} observations
\citep[e.g.\@][]{Dasadia2016a,Dasadia2016}. Thus, while the CF may be
detected, the configuration of two opposite shocks with a CF in
between, which marks the collision of an inflowing stream, will be
difficult to identify.

In \cref{fig:cl107XrayBoreBottom,fig:cl107XrayBoreTop} we show the
X-ray surface brightness profiles for the CF and shocks studied in
CL107 for the bottom and top regions (respectively) along the same
line used for the profiles in
\cref{fig:cl107BoreBottom,fig:cl107BoreTop}. The surface brightness
for these features is lower by 1--2 orders of magnitude compared to
the features in CL6, which most likely sets these features beyond
current observational capabilities. The difference is not surprising
since the features are found 2.5--3.5 times farther (in projected
distance) from the centre of the cluster. For an isothermal-model ICM
$(\rho\propto r^{-2})$ the surface brightness is expected to drop as
$S_x \propto d^{-3}$, where $d$ is the projected distance form the
cluster centre.

In this paper we have carried out an in-depth study of 2 examples of
CFs linked to collisions of inflowing streams.  It is important to
note that these examples are by no means the only instances of CFs which are 
associated with stream collision in the simulation suite we have
examined. A detailed study of all the clusters in our sample is beyond
the scope of this work, but even a cursory search reveals that CFs are
prevalent in all clusters, with many clusters hosting multiple
CFs.

We wish to make a rough estimate as to the prevalence of CFs in the
simulation suite and their detectability. To do so, we examined the
\zeq{0} snapshots of the 16-cluster simulation suite and visually
identified CFs based on maps of the hydrodynamic and thermodynamic
properties - temperature, density, pressure and entropy maps as well
as metallicity and Mach number (e.g.\@
\cref{fig:cl6VelMach,fig:cl6Maps}). If a CF was found in close
vicinity to an inflowing stream with no discernible satellite in the
region and lacking a sharp metallicity gradient, it was marked as
possibly linked to stream collisions. On average, we found
approximately 6 CFs per cluster and of these, \mbox{$\sim 50$} per
cent appear to be linked in some way to inflowing streams.

Since the X-ray flux is a strong function of the gas density, we use the
number density of the gas as a coarse estimate of detectability. In
observational studies of shocks and CFs, a typical number density of features
detectable with the \emph{Chandra} observatory is \mbox{$\sim
  10^{-3}\units{cm^{-3}}$} \citep{Markevitch2007,Dasadia2016a,Dasadia2016}. We
use this value as a detectability limit and mark CFs with local densities
above this value as possibly detectable. In total, roughly 15 per cent of all
CFs are detectable with \emph{Chandra} and of the sub-group of CFs linked to
inflowing streams, roughly 13 per cent are detectable. These fractions will
most likely be substantially larger for the next generation of X-ray missions,
\emph{Athena} \citep{Nandra2013} and the {\em X-ray Surveyor}
\citep{Weisskopf2015}.

\section{Summary and Discussion}\label{sec:discuss}
In this paper we report a robust mechanism for generating shocks and
CFs in the central regions of clusters via the inflowing gas streams,
which are seen in simulations to be prevalent in many clusters. This
mechanism should be particularly relevant in unrelaxed clusters (see
e.g. \citealt{Hallman2010}) in which gas streams are seen to penetrate
into the core \citep{Zinger2016}.

Inflowing gas streams, originating in the large scale filaments of the cosmic
web, account for most of the mass accretion into the systems and can travel at
high velocities with $\gtrsim 1000 \units{km\,s^{-1}}$, carrying with them a
significant amount of energy.  In clusters, they are heated to the virial
temperature as they penetrate through the halo. In some cases, streams are
seen to penetrate into the central regions of the cluster whereas in others,
the stream stop or dissipate before reaching the centers. The dynamical state
of the clusters was found to be linked to the degree of penetration
\citep{Zinger2016}. These penetrating streams often collide either with other
streams or with the existing ICM, and thus can lead to the formation of shocks
and CFs.

We examined an idealized 1D scenario for a collision between two
streams (or a single stream and the relaxed ambient gas) of constant
density and pressure and found that as a result of the collision, two
shocks are formed, propagating in opposite directions. Between the two
shocks a contact discontinuity in density invariably forms, which
travels at the velocity of the post-shock gas. The contact
discontinuity is in pressure equilibrium, thus a jump in temperature
is expected to be compensated by a drop in gas density, except in the
exceptional case of completely identical streams in both density and
pressure.

This simple, idealized picture is a far cry from the complex 3D
structures and processes found in the ICM in observations and in
simulations such as the ones analysed here. However, detailed
examination of stream collisions in simulated clusters revealed
configurations which resemble the idealized test-problem.

In the cluster CL6, at the site of the stream collision, two shocks
moving in opposite directions were identified with a distinct CF found
between them, whose density and temperature contrasts are (inversely)
equal to within $\lesssim 5$ per cent. The absence of any form of
substructure and the dearth of metals in the gas at the CF location
enables us to rule out satellites as a source of the CF.  An
additional shock, likely formed earlier, was found beyond the leading
shock and analysis of the shock properties suggests that the shocks
will merge and lead to the formation of another CF
\citep{Birnboim2010}.

We investigated the CF formation in a fashion similar to observations,
namely analysing a single snapshot to determine the link between the
CFs and the streams that generated them. A future study, utilizing
simulations with improved temporal resolution is planned in order to
study the formation and evolution of CFs formed by stream collision
over time.

The primary objective of this paper is to provide a proof of concept for the
formation of shocks and CFs by the collisions of inflowing gas streams
from the cosmic web. The particular clusters presented here were
chosen since it demonstrated a clear and compelling example for the
mechanism. The stream collision site was fortuitously situated in such
a way as to allow easy visualization of the CF along the Cartesian
projections of the simulation.

Examining the potential to detect such CFs in observations, we found that for
prominent cases, such as the CF found in CL6, observational detection is
definitely within the current capabilities of deep {\em Chandra} observations,
although identifying the full configuration of two oppositely oriented shocks
with a CF in between may prove challenging. In addition, one must bear in mind
that while stream collisions and subsequent formation of CFs can occur
anywhere in the cluster, the potential to detect them is highest in the
central regions where the X-ray emmission is strongest.

The cases examined in this paper are by no means the only instances of CFs
which are associated with stream collisions in the simulation suite we have
examined. A visual survey of the entire simulation suite at \zeq{0} yielded
multiple CFs in all the simulated clusters with roughly half of all CFs
showing a possible connection to the inflowing streams. In our rough
estimation, \mbox{$\sim 15$} per cent of all CFs are potentially detectable
with current instruments.

Linking CFs to streams unequivocally is only possible with an in-depth
analysis as presented in the paper, but we found CFs that resulted from
the collision of inflowing streams in nearly all clusters we examined.

One such example is presented in \rfsec{cl107}, where we examine an
additional cluster (CL107) in which colliding gas streams generate
shocks and CFs. In this cluster, the shock front at the collision site
is comprised of two distinct shocks, one originating from the stream
collision and the other from the motion of a large satellite travelling
with the stream. In the latter, the resulting CF contains a
metallicity gradient across the CF, indicating that gas stripped from
the satellite is partially responsible for the CF.

The examples brought forth in this paper highlight the challenge of
disentangling the contribution of the gas flowing along the stream and
contribution of the merging sub-structure to the formation of the
shocks and CFs. Since the gas streams mark the preferred direction of
accretion into the cluster, merging satellites are often found
travelling along the inflowing streams.  In addition, it has been
shown that the gas stripped in major mergers in the cluster can stream
towards the centre in high-velocity flows which resemble the large
scale gas streams we describe here \citep{Poole2006}.

In many of the other examples of CFs found at stream collision sites
in our simulation suite we found additional features, such as
satellites, which made it difficult to link the CF to the stream
unequivocally. To complement the findings in this paper, a study of
the prevalence of CFs at stream collision sites and their properties
is in order. In particular, it is important to ascertain how common
this mechanism is compared to other processes which form CFs.

It may be argued that since both mergers and streams are aspects of
the mass accretion, there is no point in differentiating between them
as mechanisms of CF formation. Indeed, as shown in \rfsec{coldFronts},
high-velocity streams, regardless of their origin, will lead to the
formation of shocks and CFs in the ICM.
  
However, merger events in the core are episodic by nature and their
effect on the ICM only lasts for $\sim 0.1\textrm{--}1 \units{Gyr}$
\citep{Churazov2003,Tittley2005,Ascasibar2006,Poole2006} whereas the
accretion through streams can be continuous for longer periods of
time.

Beyond the simple considerations of detectability, the ability to
observe CFs is dependent on their stability to various physical
processes which can destroy them. Thermal conduction and particle
diffusion, for example, can smear out the features of CFs on
time-scales of $\sim 10\units{Myr}$ \citep{Markevitch2007}, which
implies that in order for CFs to be observed as much as they are, they
must either be formed frequently, or that other factors, such as
magnetic fields \citep{Carilli2002}, are suppressing the thermal
conduction. Magnetic fields can also suppress Kelvin-Helmholtz
instabilities from breaking up CFs
\citep{Keshet2010b,Roediger2013,ZuHone2011,ZuHone2015}. Shocks
crossing the CF can also disrupt it via the Richtmeyer-Meshkov
instability \citep{Brouillette2002}.

An important aspect of generating CF by stream collisions is that
while individual CF may disappear, new ones will constantly form on
time-scales of several $\unitstx{Gyr}$ as long as the streams
persist. When comparing the cluster CL6 at two different epochs we
found that the inflowing gas streams persisted over several
$\unitstx{Gyrs}$, but that the penetration depths and thus the shocks
and CFs generated changed over time.

As a case in point, at the stream collision site at \zeq{0} we found
two shocks propagating upwards (red and green arrows in
\cref{fig:cl6Maps}) which are expected to merge in \mbox{$\sim
  350\units{Myr}$}. At the location where the shocks merge, a new CF
will be formed \citep{Birnboim2010}.  Another point to consider is
that formation of CFs via stream collisions is a natural explanation
for CFs found in clusters in which there is no evidence of merger
events.

In absence of instruments that can directly observe the gas streams in
clusters, identifying the shocks and CF formed at the collision site
may afford an indirect way to identify the streams. Observation of a
double shock configuration with a CF found in between, as presented
above, would constitute a strong piece of evidence for the existence
of gas streams in clusters, beyond the realm of cosmological
simulations.

\section*{Acknowledgments}
We thank Neta Zinger for his assistance and advice. We thank the
anonymous referee for comments and advice which improved the
paper. This work was partly supported by the grants ISF 124/12, I-CORE
Program of the PBC/ISF 1829/12, BSF 2014-273, and NSF AST-1405962 and
AST-1412768.

\bibliographystyle{mnras}
\bibliography{zinger_coldFront_bib}

\label{lastpage}

\end{document}